\documentclass[twocolumn, conference, comsoc]{IEEEtran}
\usepackage[noend]{algpseudocode}
\usepackage{graphicx}
\usepackage{algorithm}
\usepackage{algorithmicx}
\usepackage{color}
\usepackage[utf8x]{inputenc}
\usepackage{balance}
\usepackage[font=scriptsize]{subcaption}
\usepackage[font=scriptsize]{caption}

\IEEEoverridecommandlockouts
\begin{document}

\title{Source Localization and Tracking for Dynamic Radio Cartography  using Directional Antennas}

\author{Mohsen Joneidi$^*$,
        Hassan Yazdani$^*$,~Azadeh Vosoughi,~\IEEEmembership{\small{Senior Member,~IEEE}}
        {\small and}~Nazanin~Rahnavard,~\IEEEmembership{\small{Senior Member,~IEEE}}\\
        \IEEEauthorblockA{{Department of Electrical and Computer Engineering}\\
{University of Central Florida}\\
Emails: \{mohsen.joneidi@, h.yazdani@knights., azadeh@, and  nazanin@eecs.\}ucf.edu}
\thanks{* Indicates shared first authorship. This material is based upon work supported by the National Science Foundation under grants CCF-1718195, CCF-1341966, and ECCS-1443942.}
}
%\IEEEoverridecommandlockouts\IEEEpubid{\makebox[\columnwidth]{978-1-7281-2294-6/19/\$31.00 ©2019 IEEE \hfill} \hspace{\columnsep}\makebox[\columnwidth]{ }}
        % <-this % stops a space
%\thanks{* means shared first authorship.  M. Joneidi, H. Yazdani, Azadeh Vosoughi and N. Rahnavard are with the Department of Electrical and Computer Engineering, University of Central Florida, Orlando, Fl, 32816 USA e-mails:  mohsen.joneidi@ucf.edu, h.yazdani@knights.ucf.edu, azadeh@ucf.edu and nazanin@eecs.ucf.edu.}% <-this % stops a space
% <-this % stops a space

% The paper headers
%\markboth{Submitted to IEEE Journal on Selected Areas in Communications}%
%{Shell \MakeLowercase{\textit{et al.}}: Bare Demo of IEEEtran.cls for IEEE Journals}

\maketitle 

\begin{abstract}
 Utilization of directional antennas  is a promising solution for efficient spectrum sensing and accurate source localization and tracking. Spectrum sensors equipped with directional antennas should constantly scan the space in order to track emitting sources and discover new activities in the area of interest. In this paper, we propose a new formulation that unifies received-signal-strength (RSS) and direction of arrival (DoA) in a compressive sensing (CS) framework. The underlying CS measurement matrix is a function of beamforming vectors of sensors and is referred to as the propagation matrix. Comparing to the omni-directional antenna case, our employed propagation matrix provides more incoherent projections, an essential factor in the compressive sensing theory.
 Based on the new formulation, we optimize the antenna beams, enhance spectrum sensing efficiency, track active primary users accurately and monitor spectrum activities in an area of interest. In many practical scenarios there is no fusion center to integrate received data from spectrum sensors. We propose the distributed version of our algorithm for such cases. Experimental results show a significant improvement in source localization accuracy, compared with the scenario when sensors are equipped with omni-directional antennas.
Applicability of the proposed framework for dynamic radio cartography is shown.  
 Moreover, comparing the estimated dynamic RF map over time with the ground truth demonstrates the effectiveness of our proposed method for accurate signal estimation and recovery.
 %the estimated power map of PUs over time using the proposed method for directional antennas is plotted which shows the accuracy of estimation and signal recovery.
\end{abstract}


\IEEEpeerreviewmaketitle

%******************************************************************
\section{Introduction}
The cognitive radio (CR) paradigm is a promising solution to alleviate today's spectrum deficiency caused by an increasing demand for ubiquitous wireless access \cite{Akyildiz06_xG, Dynamic_Spectrum_Access4205091, AsilomarPaper, GlobalSIP}. In the static spectrum allocation strategies, the licensed holders of the spectrum (a.k.a. primary users or PUs) often under-utilize this valuable resource \cite{Akyildiz06_xG}.
The CR paradigm allows the unlicensed or secondary users (SUs) to coexist with PUs and to access the spectrum as long as they do not interfere with PUs. The under-utilized spectrum bands that can be used by the SUs are called \emph{spectrum holes}~\cite{Haykin05_brain}. One approach to monitoring (sensing) spectrum and identifying the spectrum holes is through the localization of emission sources (PUs). Given the locations of PUs and the signal propagation parameters one can estimate spectrum power of an area of interest. 
%
%{\color{red}The text on centralized versus distributed location should be moved here, before talking about TDoA, RSS, DoA.}
Collaborative source localization techniques are divided into two main categories, i) centralized techniques, and ii) distributed techniques. Centralized techniques are based on the assumption that signals from all SUs are collected at a fusion center (FC), that is tasked with fusing the received signals and localizing the sources. On the other hand, distributed techniques are based on the assumption that there is no FC. In this case, each SU receives signals from its neighbors and performs source localization locally \cite{Shirazi}. 

\par Source localization can generally be performed  by estimating time-difference of arrival (TDoA), received-signal-strength (RSS), or direction of arrival (DoA). The first approach implies synchronization of SUs, which is infeasible in many scenarios. Most of the previous works on spectrum sensing exploit omni-directional antennas for SUs \cite{Bazerque10Distributed,bazerque2011group}. Thus, DoA cannot be used for source localization in these works. 
The most popular DoA estimation methods include MUSIC, Capon and ESPRIT \cite{BF1}. However, the performance of these methods are limited in low SNRs or when the sources are placed close to each other. In \cite{Cabric, Cabric2} the RSS and DoA of a PU is estimated using energy measurements from a sectorized antenna and the performance and theoretical bounds of DoA/RSS estimation are studied. The orientation of a PU with respect to SU's location is determined in \cite{J1, CISS2019} based on the RSS where an SU is equipped with a reconfigurable antenna. The signal received at an antenna array is sparse in the spatial domain and compressive sensing can be employed for DoA estimation using less number of radio frequency (RF) chains \cite{DoA_CS}. In \cite{MILCOM} the problem of DoA estimation using electronically steerable parasitic array radiator (ESPAR) antenna based on compressive sensing is exploited for a sensor  equipped with a single RF chain. However, in the present paper directional antennas are employed and their scanning pattern are updated adaptively. Moreover,  compressive sensing formulation is employed for improving the accuracy of source parameters estimation.

\par In the present paper, we assume some sensors (or SUs) are randomly deployed in the area of interest to sense spectrum  and the goal is to estimate locations of PUs and generating power spectrum map for the whole area of interest. Each sensor is equipped with an antenna with uniform linear array (ULA). The weighting (beamforming) vector corresponding to each ULA can be adapted/updated to steer/direct the antenna beam and thus localize PUs within the area and track them accurately (if they are mobile). We develop both centralized and distributed collaborative source localization algorithms, in which RSS and DoA information are integrated in a unified formulation. 
%Moreover, weighting vector of collaborative antennas are updated efficiently in order to track active PUs and cover spectrum activities of the area of interest. 
In our formulation, we assume an area of interest is divided into $P$ grid points. PUs (emitting sources) may only reside in these grid point (Figure~\ref{fig:setup}). PUs locations and powers are modeled by a sparse vector, in which a non-zero element indicates the presence of a PU in the corresponding grid point.
%PUs (to be localized and tracked within the area) are modeled as a sparse vector where the grid points in an area correspond to the elements of the vector and non-zero elements in the vector indicate the presence of a PU in the corresponding grid point. 
The received signal at each SU is a superposition of the emitted signals by sources (whose locations and signal powers are unknown). In the centralized setup, we establish a compressive sensing problem at the FC, where the compression operator is modeled in terms of the signal propagation parameters and the sensing patterns of the SUs directional antennas (encompassing ULA structure and DoA information as well as weighting/beamforming vectors).  In the distributed setup,  each SU shares its received signal with its neighbors (determined by the connectivity graphs), and hence each SU can establish a compressive sensing problem, based on the collected received signals from its neighbors. For both centralized and distributed setups, we provide source localization and tracking algorithms that exploit the inherent sparsity and adaptive beamforming to accomplish accurate source localization and tracking within the area. The main contributions of our paper are summarized as follows:
\begin{itemize}
    \item RSS and DoA information are integrated in a unified formulation based on compressive sensing. 
    \item Sensing patterns (or weighting vectors) of directional antennas at SUs are optimized, in order to quickly and accurately discover the PUs' activities and efficiently track them if they move within the area.
    \item In the distributed setup, neighboring SUs localize sources collaboratively and track them via adapting their weighting vectors and steering/directing their beams.
    \item As  a by-product we can estimate the power spectrum map of the area (so-called radio cartography).
    %\item The estimated power map of PUs over time using the proposed method for directional antennas is plotted which shows the accuracy of estimation and signal recovery.
\end{itemize}

\textbf{Notation}: Throughout this paper, vectors and matrices are written as bold lowercase and uppercase letters, respectively. Norm $\ell_2$ of a vector is denoted by $\|.\|_2$ and $t$ {\it{mod}} $B$ is equal to the remainder of the division $t/B$.%{\color{red}***is mod(,) a common notation for the remainder?**}.

\par The rest of the paper is organized as follows. Section \ref{sec:problem_state} states the problem and explains the employed system model. 
Sections \ref{sec:proposed} and \ref{sec:implementation} present our proposed centralized and distributed source localization and tracking algorithms.
%Section \ref{sec:implementation} describes the distributed implementation of our proposed algorithm. 
Section \ref{sec:cartography} elaborates how we can leverage on our source estimation findings in the previous sections to obtain the power map of the entire area. Section \ref{sec:experiments} exhibits our experimental results.
% ************************************************************************************
\section{System Model and Problem Statement}
\label{sec:problem_state}
\begin{figure}[t]
\hspace{3mm}
\centering
\begin{subfigure}{1.9in}
\centering
%\hspace{3mm}
\includegraphics[width=2.3in,height=1.9in]{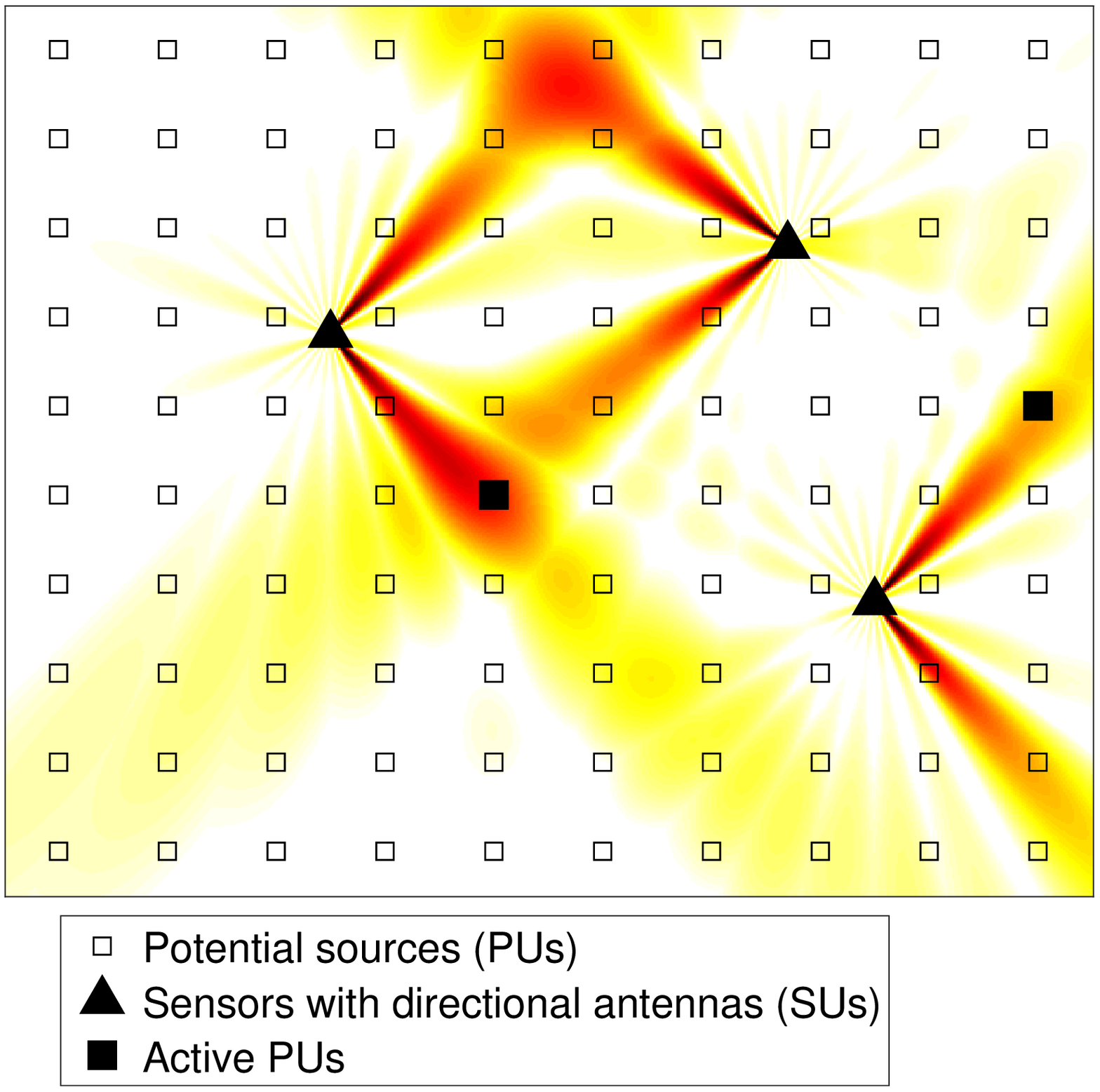}
\end{subfigure}
\begin{subfigure}{1in}
\centering
\vspace{-4mm}
\includegraphics[width=0.35in,height=1.6in]{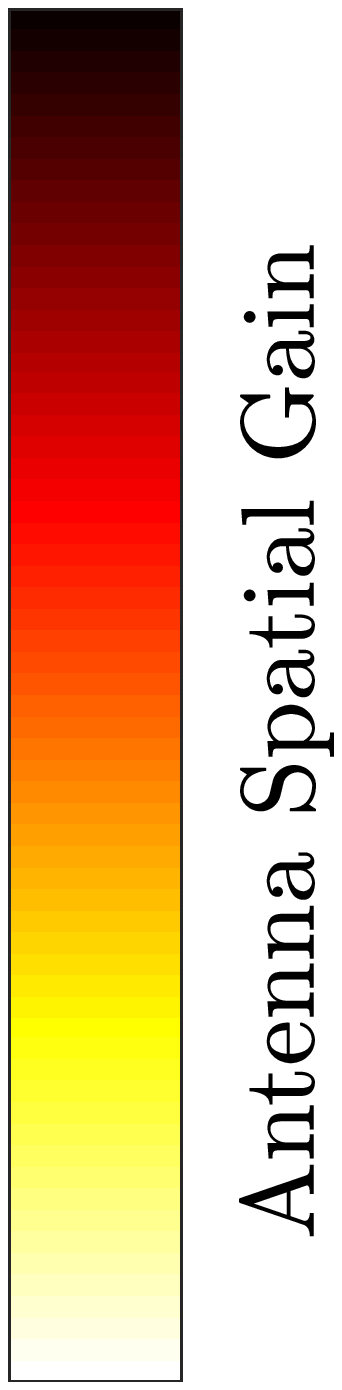}
\end{subfigure}
\caption{\small{The grid of $P=100$ grid points. The PUs are active in $2$ locations out of $100$ grid points shown by filled black squares and  $3$ SUs sense the environment using their directional antennas. The goal is to localize and track the sources and to estimate a dynamic power map, leveraging on SUs' capabilities of adaptive beamforming.}}
\label{fig:setup}
\vspace{-3mm}
\end{figure}

Our system model can be viewed as the extension of the system model in~\cite{Bazerque10Distributed}, for the case when SUs are equipped with \emph{directional antennas} (instead of omni-directional antennas). Suppose an area of interest is divided into $P$ grid points as shown in Fig. \ref{fig:setup} and PU transmitters are assumed to be located in a subset of these grid points, unknown to SUs (i.e., the number and the locations of PU transmitters as well as their transmission power are unknown to SUs). Moreover, there are $N$ SUs (or spectrum sensors) with known locations, where each sensor receives a superposition of the PUs' signals subject to a zero mean measurement noise.  We suppose antenna of each SU is a ULA consisting of $M$ elements  with adjacent element spacing of $d$.  Assuming $\theta_{n,p}$ is the orientation of grid point $p$ with respect to sensor $n$, the manifold matrix for sensor $n$ is given by $\boldsymbol{A}_n= [a_{n,m,p}]$, where 
%
% ---------------------------------------------------------------
\begin{align}\label{anmp}
a_{n,m,p}= \frac{1}{R_{n,p}^\eta} e^{-j\frac{2\pi d(m-1)}{\lambda_c} sin(\theta_{n,p})},
\end{align}
% ---------------------------------------------------------------
%

in which $\lambda_c$ is the wave-length corresponding to the operating frequency of each SU. Moreover, $R_{n,p}$ is the distance between grid point $p$ and sensor $n$, and $\eta$ is the path-loss exponent \cite{BF1}. We note that in general, the received signal at sensor $n$ would be affected by channel fading gains between PU transmitters and sensor $n$. These gains can be obtained by training sequences, which is beyond the scope of this paper. For simplicity, similar to \cite{Bazerque10Distributed,bazerque2011group}, we only consider path loss effect in our signal modeling in (\ref{anmp}), to indicate that the received signal will be attenuated inversely proportional to $R_{n,p}$\footnote{The propagation model can be extended for fading channels by assuming an uncertainty on the channel gains.}. We write the received signal at sensor $n$ as 
%
% ---------------------------------------------------------------
\begin{align}\label{eq:rn}
r_n= \boldsymbol{w}_n^H \boldsymbol{A}_n \boldsymbol{s} + \boldsymbol{w}_n^H \boldsymbol{z}_n,
\end{align}
% ---------------------------------------------------------------
%
where $\boldsymbol{w}_n = [w_{n,1}\;  \ldots\; w_{n,M}]^T$ is the weighting (or beamforming) vector at sensor $n$, and we assume that $\| \boldsymbol{w}_n \|_2^2=C_1, \forall n$ where $C_1$ is the antenna gain. In \eqref{eq:rn} $\boldsymbol{s} = [s_1\; \ldots\; s_P]^T$ is the transmitted signal vector from PUs, and $\boldsymbol{z}_n = [z_{n,1} \; \ldots\; z_{n,M}]^T$ is the received noise vector at sensor $n$ corresponding to its ULA, where $z_{n,m} \sim {\cal C}{\cal N}(0,\sigma_{z_n}^2)$ are zero-mean complex Gaussian independent and identically distributed (i.i.d). {Fig. \ref{fig:setup} illustrates an example of antenna spatial gain (or sensing pattern) of sensors at a single time slot. However, the sensing pattern can be a different one at the next time slot. The beam of antennas in a set of  consecutive time slots should (i) cover the whole area in order to discover a new appearing PU and (ii) scan the location of previously estimated active PUs to track their changes. }

We assume that the signals of PUs' transmitters  are independent and  let $y_n (t)=|r_n(t)|^2= r_n(t) r^*_n(t)$ denote the RSS of sensor $n$  at time $t$. We can write $y_n (t)$ as the following:
%\vspace{-1mm}
%\gamma_{pn} x_p(t)+v_n(t)
%
%---------------------------------------------------------
\begin{align}
\label{eq:RSS}
%\begin{eqnarray}
%\begin{split}
y_n (t) = &\sum_{p=1}^P \underbrace{ \boldsymbol{w}_n^H(t) \boldsymbol{A}_n \boldsymbol{C}_p \boldsymbol{A}_n^H\boldsymbol{w}_n(t)}_{=\gamma_{pn}(t)} x_p(t) \nonumber \\
\scriptsize{\;\;} \nonumber \\
 + & \underbrace{ \boldsymbol{w}_n^H(t) \boldsymbol{z}_n(t) \boldsymbol{z}_n^H(t) \boldsymbol{w}_n(t)}_{=v_n(t)} =\boldsymbol{\gamma}_n^T(t) \boldsymbol{x}(t) +v_n(t),
%\end{split}
%\end{eqnarray}
\end{align}
%---------------------------------------------------------
where the  vector $\small{\boldsymbol{\gamma}_n(t) =[\boldsymbol{\gamma}_{1n}(t)\; \;\boldsymbol {\gamma}_{2n}(t)\; \;\ldots\;  \boldsymbol {\gamma}_{P n}(t)]^T}$, and
$x_p(t)=|s_p(t)|^2$ represents the propagation power at the grid point $p$ and time instant  $t$, $\boldsymbol{x}(t)=[x_1(t)\; x_2(t) \; \ldots \;  x_P(t)]^T$ is the propagation power vector, $\boldsymbol{w}_n(t)$ is the beamforming vector that sensor $n$ uses at time $t$, and $v_n(t)$ is distributed as an exponential random variable with mean $m_{v_n} = \| \boldsymbol{w}_n \|_2^2 \sigma^2_{z_n}=C_1 \sigma^2_{z_n}$.
Matrix $\boldsymbol{C}_p$ is a $P\times P$ matrix with all entries equal to $0$ except the diagonal entry of index $(p,p)$. Note that $\gamma_{n,p}$ includes the pathloss effect as well as the beamforming vector $\boldsymbol{w}_n$.  We can write $v_n(t)$ in \eqref{eq:RSS} as
%
%---------------------------------------------------------
\begin{align}
v_n(t) = \varepsilon_n(t) + m_{v_n},
\end{align}
%---------------------------------------------------------
%
where $\varepsilon_n(t)$ is a zero-mean random variable with variance $m_{v_n}^2$. We assume that sensor $n$ can adjust its antenna sensing pattern and steer its beam, via optimizing its beamforming vector $\boldsymbol{w}_n$. Examining \eqref{eq:RSS} we note that vector $\gamma_n$ depends on the optimization variable $\boldsymbol {w}_n$. Moreover, vector $\boldsymbol{x}(t)$ is unknown and sparse. Assuming the RSS $y_n(t)$ for some $n$ values (in the centralized setup, $n \in \{1, \ldots ,N\}$, and for the distributed setup, $n \in {\cal N}_n$,  where ${\cal N}_n$ is the set of neighboring sensors of sensor $n$) are available, our main goal is to exploit the steering capability of SUs' antennas in order to improve the accuracy and efficiency of spectrum sensing, for the general scenario when the number and the locations of PU transmitters are unknown and time varying. This means that we aim at finding an accurate estimate of the sparse vector $\boldsymbol{x}(t)$ as time changes, as we allow SU sensors to adjust their antenna sensing patterns. By allowing SUs to steer their beams and optimize their beamforming vectors, we enable discovering of new PU signals and tracking those that have been discovered previously. Having an accurate estimate of $\boldsymbol{x}(t)$ we can build the dynamic radio power map of the field, via estimating the RSS at each grid point and time instant $t$. We note that as we solely use RSS for estimating $\boldsymbol{x}(t)$ and the power map, no synchronization is required among SU sensors.

\par Considering centralized estimation of $\boldsymbol{x}(t)$, let $\boldsymbol {y}(t)=[y_1(t)\;y_2(t)\;\ldots \; y_N(t)]^T$ denote the vector containing the RSS of all $N$ sensors. Moreover, let $\boldsymbol{\Gamma}$ denote an $N \times P$ matrix, whose $n$-th row is $\boldsymbol{\gamma}_n^T$. Matrix $\boldsymbol{\Gamma}$ is referred to as the measurement matrix in the literature of compressive sensing and we refer to it as propagation matrix here. The vector of RSS at the FC can be written as
% ---------------------------------------------------------------
\begin{equation}
\boldsymbol{y}(t)=\boldsymbol{\Gamma}(t)\boldsymbol{x}(t)+\boldsymbol{m}_v + \boldsymbol{\varepsilon}(t),
\end{equation}
% ---------------------------------------------------------------
where $\boldsymbol{m}_v=[ m_{v_1}\; m_{v_2}\; \dots\; m_{v_N}]^T$ is the vector of averaged RSS and vector $\boldsymbol{\varepsilon}(t)$ is the zero-mean residual of all sensors at time instant $t$ . The centralized estimation of the sparse signal $\boldsymbol{x}(t)$ can be viewed as the solution to the following constrained optimization problem
\begin{align}
\label{eq:main_problem1}
<\hat{\boldsymbol{x}} (t),\hat{\boldsymbol{W}} (t)>&=\underset{\boldsymbol{x},\boldsymbol{W}}{\text{argmin}} \|\boldsymbol{y} (t)-\boldsymbol{\Gamma}(t)\boldsymbol{x}-\boldsymbol{m}_v\|^2_2\nonumber \\ 
\text{s.t.} \; \boldsymbol{\Gamma}(t)=h(\boldsymbol{W}(t))&,\; \; \|{\boldsymbol{x}}\|_1\le {C_0 }\;  \text{and} \; \|{ {\boldsymbol{w}}_n} \|_2^2 =C_1 ~\forall n,
\end{align}
% ---------------------------------------------------------------
%
where $\boldsymbol{W}= [\boldsymbol{w}_1\; \ldots \; \boldsymbol{w}_N]$ is an $M \times N$ matrix containing beamforming vectors and $h(\cdot)$ is a function of beamforming matrix $\boldsymbol{W}$ which can be derived using \eqref{eq:RSS}. %Also, { \color{blue}  $\boldsymbol{m}_v$ is the average noise  of all SUs}.
Parameter $C_0$ tunes the impact of $\ell_1$ regularization and parameter $C_1$ restricts antenna gain  of SUs. Fig. \ref{fig:setup} shows a sensor network with $3$ SUs and $100$ candidate points for PUs where only $3$ of them are active and propagate signals. SUs continuously scan the space to discover signals of new PUs and track the previously detected ones. 
\par Previous works have focused on estimation of variable $\boldsymbol{x}$ subject to a given propagation matrix. The following  minimization problem has been proposed for estimating  $\boldsymbol {x}(t)$ for each time independently \cite{Bazerque10Distributed,joneidi2017dynamic}:
%
% ---------------------------------------------------------------
\begin{equation}
    \label{eq:l1}
\hat{\boldsymbol{x}} (t)=\underset{\boldsymbol{x}}{\text{argmin}} \| \boldsymbol{y} (t)-\boldsymbol{\Gamma}\boldsymbol{x}-\boldsymbol{m}_v\|_2^2+\lambda \|{\boldsymbol{x}}\|_1. 
\end{equation}

In Problem (\ref{eq:l1}), propagation matrix is a function of distance of SUs and PUs and it can not be optimized. However, in the present work we employ directional antennas and beamforming is needed to track existing PUs and discover any appearing PUs.
% ---------------------------------------------------------------
 %{\color{red}Problem (\ref{eq:l1}), subjected to $\ell_1$  regularization, has previously been employed for collaborative spectrum estimation for a given $\boldsymbol{\Gamma}$ \cite{Bazerque10Distributed,joneidi2017dynamic}.}  {\color{red}
 Next, we will discuss the centralized and distributed solutions of our constrained optimization problem. 
% ---------------------------------------------------------------

% ---------------------------------------------------------------
%
%
%**********************************************************************
%**********************************************************************
\section{Centralized Source Localization and Tracking}
\label{sec:proposed}
In this section, we address Problem \eqref{eq:main_problem1}. To achieve this, we need to link between the entries of matrix $\boldsymbol{\Gamma}$ and dynamic of network through beamforming matrix $\boldsymbol{W}$, and the last estimate of the propagation power vector $\boldsymbol{x}(t\!-\!1)$. Each SU sweeps a different pattern at each time slot. Assume at time slot $t$ we jointly estimate the propagation power vector $\boldsymbol{x}(t)$, and optimize the beamforming vector $\boldsymbol{w}_n(t)$. We exploit $B$ recent measurements from $B$ distinguished sensing patterns. Each $B$ consecutive time slots make a time block of size $B$. Let us break Problem \eqref{eq:main_problem1} into two alternating subproblems as the following
% ---------------------------------------------------------------
\begin{subequations}
\label{eq:break}
\begin{align}
\label{eq:break1}
\hat{\boldsymbol{x}} (t)&=\underset{\boldsymbol{x}}{\text{argmin}} \|\boldsymbol{y}^B (t)-\boldsymbol{\Gamma}^B(t)\hat{\boldsymbol{x}}-\boldsymbol{m}_v^B\|_2^2 +\lambda \|\boldsymbol{x}\|_1,\\
\label{eq:break2}
\hat{\boldsymbol{W}} (t\!+\!1)&=\underset{\boldsymbol{W}}{\text{argmin}} \|\boldsymbol{y}^B (t)-\boldsymbol{\Gamma}^B(t\!+\!1)\hat{\boldsymbol{x}}(t)-\boldsymbol{m}_v^B\|_2^2, \\
\text{s.t.} &\; \boldsymbol{\Gamma}^B(t\!+\!1)=h(\boldsymbol{W}(t)).\nonumber
\end{align}
\end{subequations}
% ---------------------------------------------------------------
%
The first subproblem estimates $\boldsymbol{x}(t)$ using $B$ recent measurements of $N$ sensors and the second subproblem updates the next beamforming vector for the next time slot. In \eqref{eq:break1}, $\boldsymbol{y}^B(t)\! =\! \big [\boldsymbol{y}^T\!(t\!-\!B\!+\!1)\; \boldsymbol{ y}^T\!(t\! -\!B\!+\!2)\; \ldots \; \boldsymbol{ y}^T\!(t) \big ]^T$, is a vector containing  $NB$ sensed RSS measurements and $\boldsymbol{\Gamma}^B\!(t) = [\boldsymbol {\Gamma}^T\!(t \!-\!B\!+\!1)\; \boldsymbol{ \Gamma}^T \!(t\! -\!B\!+\!2)\; \ldots \; \boldsymbol{ \Gamma} ^T\!(t) \big ]^T$ is an $NB\times P$ matrix. Note that in each time slot $N$ new beamforming vectors are updated for $N$ sensors. In other words, $N$ new rows are added to matrix $\boldsymbol{\Gamma}^B$ and $N$ obsolete rows are removed.
Moreover, in $\boldsymbol{\Gamma}^B(t\!+\!1)$  the first $(B\!-\!1)\times N$ rows are shared with $\boldsymbol{\Gamma}^B(t)$. However, the last $N$ rows of $\boldsymbol{\Gamma}^B(t\!+\!1)$ are subject to optimization. %The function $h(\cdot)$ can be derived using \eqref{eq:RSS}. 
In the rest of this section we explain how to solve this problem.
\par Solving Problem \eqref{eq:break1} w.r.t. $\boldsymbol{x}$ is a straightforward problem which is discussed in the context of sparse regression and compressive sensing. However, Problem \eqref{eq:break2} is a challenging problem and the main contribution of this paper is solving \eqref{eq:break2} efficiently, i.e.,  updating beamforming vectors efficiently. 
%
%********************************************************************
\subsection{Recovery of sparse sources}
Subproblem (\ref{eq:break1}) is a classic regression problem. Usually number of sensors ($N$) are much smaller than the number of grid points ($P$), which results in a compressive sensing problem. $\ell_1$ regularization promotes sparsity that plays a key role for sparse recovery. There have been extensive efforts for solving this problem efficiently. Some of the state-of-the-art algorithms are least absolute shrinkage and selection operator (LASSO) \cite{bazerque2011group} and iterative re-weighted least squares (IRLS) \cite{MILCOM,zaeemzadeh2017robust}. There are two essential factors in compressive sensing theory (i) sparsity of the underlying process and (ii) incoherent projections from the process \cite{candes2007sparsity}. In the next subsection we study the second factor in order to provide a set of incoherent projections of $\boldsymbol{x}$ in (\ref{eq:break}).    

%********************************************************************
\subsection{Successive optimization of beamforming vectors}
Problem (\ref{eq:break2}) can be regarded as updating matrix $\boldsymbol{\Gamma}$ in order to project the most information from $\boldsymbol{x}$ into $\boldsymbol{y}$ in each time slot. There are two phases for updating entries of $\boldsymbol{\Gamma}$: phase i)  A part of $\boldsymbol{\Gamma}$ is optimized for sensing a general unknown $\boldsymbol{x}$ (discovering), phase  ii) A part of $\boldsymbol{\Gamma}$ is allocated and optimized for tracking a known previously estimated $\boldsymbol{x}$ (tracking). In the first phase unseen spectrum activities are discovered. While in the second phase an estimate for $\boldsymbol{x}$ is available and small changes are tracked via optimizing beamforming vectors of sensors. 

%\subsubsection{Discovery of unknown sources}
$\bullet$ \textbf{Phase 1:} {\it Discovery of unknown sources}:
First we write the problem in terms of optimization of $\boldsymbol{\Gamma}$. 
% ---------------------------------------------------------------
\begin{equation}
    \label{eq:gamma_opt1}
\boldsymbol{\Gamma}=\underset{\boldsymbol{\Gamma}}{\text{argmin}} \;\mathbb{E}_x \big \{\|\boldsymbol{y} (t)-\boldsymbol{m}_v^B\;-\boldsymbol{\Gamma}{\boldsymbol{x}}\|_2^2 \big \}.
\end{equation}
% ---------------------------------------------------------------
%
In the case of Gaussian and independent noise for each sensor it is easy to show that this problem is equal to solving the following problem,
%
% ---------------------------------------------------------------
\begin{equation}
        \label{eq:gamma_opt2}
\boldsymbol{\Gamma}=\underset{\boldsymbol{\Gamma}}{\text{argmin}}\; \text{trace}\Big(\sum_n \boldsymbol{\gamma}_n\boldsymbol{\gamma}_n^T \Big)^{-1}.
\end{equation}
% ---------------------------------------------------------------
%
This problem enforces rows of $\boldsymbol{\Gamma}$ to be linearly independent. In the case of linearly dependent set of rows of $\boldsymbol{\Gamma}$ the cost function in \eqref{eq:gamma_opt2} will be infinity. Intuitively, as we measure 
less correlated projections of an unknown vector we can recover it more accurately\footnote{We say two vectors are less correlated when their inner vector product has a smaller absolute value.}. Equation \eqref{eq:gamma_opt2} indicates that we should regularize Problem \eqref{eq:break2} accordingly. On the other hand, Problem \eqref{eq:break2} is expressed in terms of the beamforming matrix $\boldsymbol{W}$.
\noindent According to \eqref{eq:RSS}, row space of $\boldsymbol{\Gamma}$ is spanned by $\boldsymbol{w}_n^H\boldsymbol{A}_n$ for all $n$. Problem (\ref{eq:gamma_opt2}) should be cast in terms of these bases. Let  $\boldsymbol{u}_n^T(t) = \boldsymbol {w}_n(t)^H\boldsymbol{A}_n$ be an auxiliary row vector and  matrix $\boldsymbol{U}(t)$ be a matrix whose $N$ rows are $\boldsymbol{u}_n^T(t)$ for $n=1,\ldots ,N$. 
%{\color{red}Mohsen J.: are the rows of $U$ $u_n(t)$???}
The beam update rule, inspired by  (\ref{eq:gamma_opt2}), is formulated as 
% ---------------------------------------------------------------
\begin{align}
\label{eq:w_update_reg1}
\boldsymbol{u}_n (t+1)&=\underset{\boldsymbol{u}}{\text{argmin}}  \;\text{det}(\boldsymbol{U}_B\boldsymbol{U}_B^T)^{-1} ~~~\text{s.t.} ~\|\boldsymbol{u}\|_2^2=1,
\end{align}
% ---------------------------------------------------------------
where $\boldsymbol{U}_B$ is the concatenation of all  scanned beams in the recent $B-1$ time slots and a new row $\boldsymbol{u}$.  
In this problem, $\boldsymbol{u}$ is the last row of $\boldsymbol{U}_B$ which is the subject of optimization. This problem  will be repeated $N$ times to find a new scanning beam for all $N$ sensors. Problem (\ref{eq:gamma_opt2}) is referred as A-optimal solution in the literature of optimization \cite{Boyd:2004:CO:993483}. However, (\ref{eq:w_update_reg1}) is modified to use D-optimality by minimizing determinant \cite{Boyd:2004:CO:993483}. D-optimality provides attractive properties which is helpful for efficient optimization.  D-optimal optimization is a \emph{sub-modular} problem which can be solved optimally in a greedy manner. In other words, at time slot $t$ when the beams of  $N$ SUs is updated, we can use greedy approach to update each beam independently and add the new found beam to $\boldsymbol{U}_B$. In (\ref{eq:w_update_reg1}), the collection of $\boldsymbol{u}_n$ vectors in $\boldsymbol{U}_B$ constructs a polygonal in $\mathbb{R}^P$ in which each vertex corresponds to a previously scanned beam. The proposed optimization problem finds a new vertex on the unit sphere such that the volume of the new polygonal is maximized. This criterion has many applications in sensor selection, sensor placement,  and data reduction \cite{Sen_sel_boyd,krause2008near, li2017polynomial}. We refer to the solution of this problem as the discovery-based beamforming vectors. These beamforming vectors are uncorrelated to each other and each one tries to discover a new direction which is not covered by the span of other beams. Finally, $\boldsymbol{w}_n(t\!+\!1)$ must be computed using estimated $\boldsymbol{u}_n(t\!+\!1)$ in (\ref{eq:w_update_reg1}) as follows.
% ---------------------------------------------------------------
\begin{subequations}
\label{eq:w_update_phase1}
\begin{align}
\label{eq:w_update2}
    \tilde{\boldsymbol{w}}_n(t+1)&=(\boldsymbol{A}_n\boldsymbol{A}_n^T)^{-1}\boldsymbol{A}_n\boldsymbol{u}_n(t\!+\!1),\\
    \hat{\boldsymbol{w}}_n(t+1)&=\sqrt{C_1} \; \tilde{ \boldsymbol{w} }_n (t\!+ \!1)/\| \tilde{\boldsymbol{w}}_n(t\!+\!1)\|_2.
    \end{align}
\end{subequations}
% ---------------------------------------------------------------
The number of grid points for potential PUs ($P$) should be greater than $M$ to have an invertible $\boldsymbol{A}_n\boldsymbol{A}_n^T$.  The solution to  \eqref{eq:w_update_phase1} provides a set of incoherent vectors as the measurement matrix in the compressive sensing formulation. Incoherency of measurement vectors plays a key role in compressive sensing of a general unknown $\boldsymbol{x}$ \cite{candes2007sparsity}. However, the current estimation of $\boldsymbol{x}$ gives us valuable information about eventful parts of the area of interest which is discussed next. 

$\bullet$ \textbf{Phase 2:} {\it Tracking of  sources with a priori knowledge}:
We can obtain an initial (rough) estimate of propagation power vector $\boldsymbol{x}$ using any randomly chosen set of beamforming vectors. This rough estimate of $\boldsymbol{x}$ is considered as the initial result of source localization and is used for tracking the discovered sources more efficiently at the next time slot. To achieve this, it is sufficient to update $\boldsymbol{w}_n$ according to the equations given by
% ---------------------------------------------------------------
\begin{subequations}
\label{eq:w_track}
\begin{align}
    \tilde{\boldsymbol{w}}_n(t+1)&=\boldsymbol{A}_n \hat{\boldsymbol{x}}(t), \\
        \hat{\boldsymbol{w}}_n(t+1)&=\sqrt{C_1}\tilde{\boldsymbol{w}}_n(t\!+\!1)/\|\tilde{\boldsymbol{w}}_n(t\!+\!1)\|_2.
        \end{align}
\end{subequations}
% ---------------------------------------------------------------
%

\begin{figure}
\centering
\includegraphics[width=3.45in,height=2in]{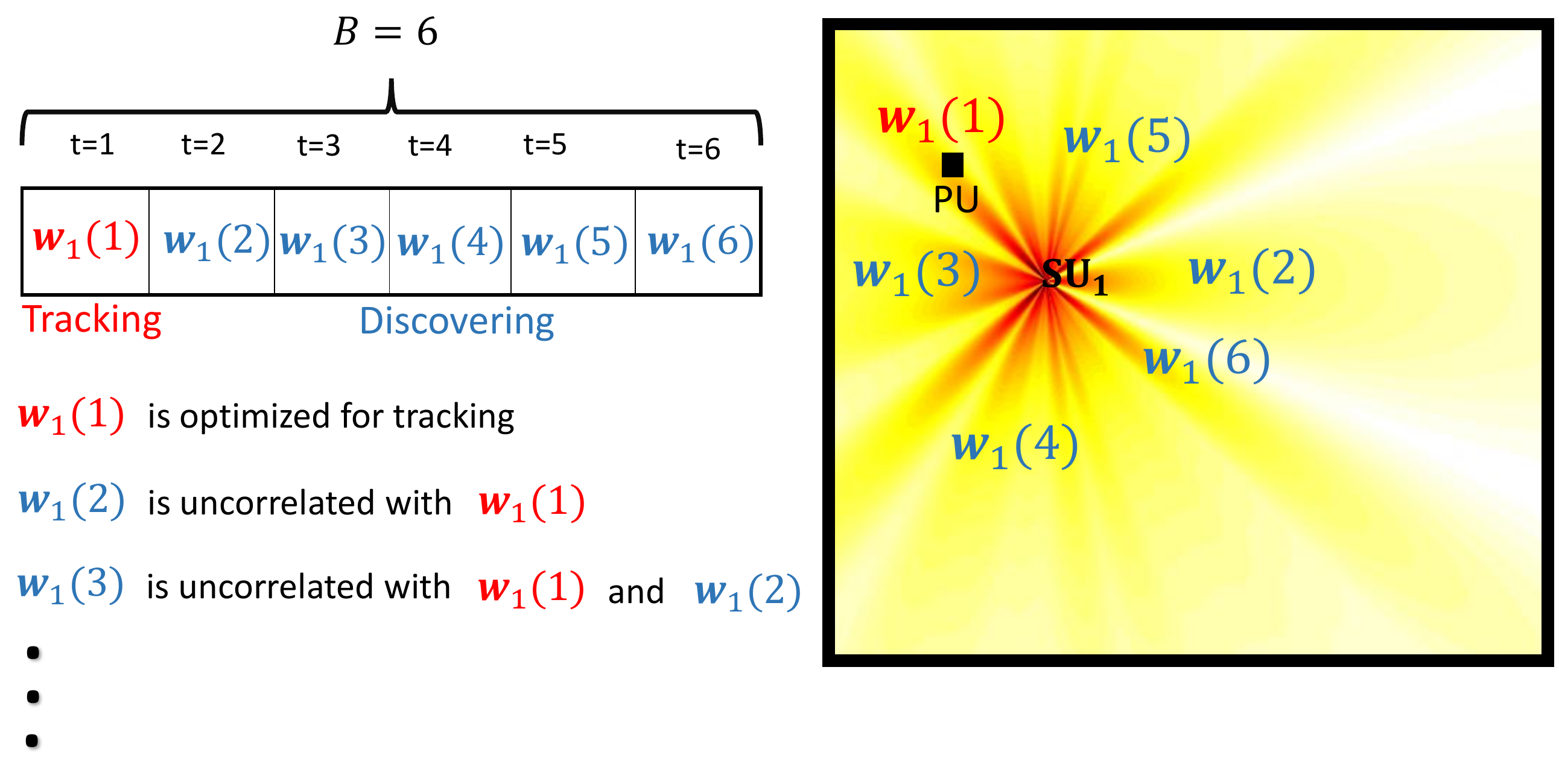}
\caption{\small{A time block of sensing operation which contains $6$ time slots. In each time block one beam is dedicated for tracking and the rest discover network's area for upcoming PUs. The main problem introduced in \eqref{eq:break} is solved using recent $B$ time slots.  }}\label{fig:time_beams}
\vspace{-4mm}
\end{figure}

This update rule performs similar to a matched filter (matched to the received signal) which maximizes the  RSS.
In each time block of operation, i.e., $B$ time slots, one beam is dedicated to be matched with the dynamic of network. The task of $B\!-\!1$ beams is to discover unseen areas of network to discover any new activity. Once a new activity is detected by estimation of $\boldsymbol{x}$, it will be refined and tracked via the dedicated beam. Assuming $NB\le M$, then discovery-based beams can be updated via selection from the null space of  beams in the recent $B\!-\!1$ time slots. Null space of these beams associates with a space that is not covered yet and the next beam is restricted to be selected from this space.  Fig. \ref{fig:time_beams} illustrates a time block of sensing operation. Assume we have a rough estimation of propagation power vector $\boldsymbol{x}$ and at the first time slot a sensing beam is optimized such that the previously detected source can be tracked. The second beam (at the second time slot) is a vector which is orthogonal to $\boldsymbol{w}_1(1)$ and at the third time slot $\boldsymbol{w}_1(3)$ is optimized to be orthogonal to the two-dimensional subspace spanned by $\boldsymbol{w}_1(1)$ and $\boldsymbol{w}_1(2)$. In general, we use the recent $B$ beams for estimation in each time. In any $B$ consecutive time slots there is one beam for tracking and $B\!-\!1$ orthogonal/uncorrelated beams\footnote{In the case of $NB\le M$ the optimum beams will be orthogonal and otherwise we can just optimize them to be uncorrelated.}. Each of these beams is orthogonal to the recent tracking beam and orthogonal to the rest of beams. 
%
% ---------------------------------------------------------------
\begin{algorithm}[b]
\caption{Centralized spectrum sensing and beamforming.}
\textbf{Input}: Location of SUs and RSS.\vspace{1mm}\\
\textbf{Output}: $\boldsymbol{x}(t)$ (Location and powers of PUs for each $t$). \vspace{1mm}\\
\vspace{1.6mm}
\hspace{-1mm} \small{1:}  Random initialization of $\boldsymbol{w}_n(t)\;\forall n$.\\
\vspace{1.6mm}
\hspace{-1mm} \small{2:}  Construct $\boldsymbol{\Gamma}$ using Eq. (\ref{eq:RSS}).\\
\vspace{1.2mm}
\textbf{for} a new time slot $t$\\
 \vspace{1mm}
 \hspace{-1mm} \small{3:}  $\qquad$ Collect vector $\boldsymbol{y}^B$ from recent $B$ time slots\\ \vspace{2mm}
   \hspace{-1mm} \small{4:} $\qquad$ Construct matrix $\boldsymbol{\Gamma}^B$ for recent $B$ time slots\\ \vspace{2mm}
   \hspace{-1mm} \small{5:}  $\qquad$ $\hat{\boldsymbol{x}} \leftarrow$ Solve Problem (\ref{eq:break1})\\\vspace{2mm}
  \hspace{-1mm}   $\;\;\;\;\qquad$ \textbf{if} $t$ {\it{mod}} $B=0$ \\ \vspace{2mm}
  \hspace{-1mm}   \small{6:}   $~~\quad\qquad$ $\hat{ \boldsymbol{w}}_n (t\!+\!1) \leftarrow$ Eq. (\ref{eq:w_track}).\\\vspace{2mm}
  \hspace{-1mm}   $\;\;\;\;\qquad$ \textbf{else} \\\vspace{2mm}
  \hspace{-1mm}   \small{7:}   $~~\quad\qquad$ Construct matrix $\boldsymbol{U}_B$ for recent $B$ time slots.\\\vspace{2mm}
  \hspace{-1mm}   \small{8:}   $~~\quad\qquad$ $\hat{ \boldsymbol{w}}_n(t\!+\!1) \leftarrow$ Problem (\ref{eq:w_update_reg1}) and Eq. (\ref{eq:w_update2}).\\
  \hspace{-1mm}   $~~~\quad\qquad$ \textbf{end} \\
$\;$\textbf{end}
\label{alg:centralized}
\end{algorithm}
% ---------------------------------------------------------------
%
In the case that $N B>M$ after $M$ time slots the null space will be empty and $\boldsymbol{w}(M\!+\!1)$ can not be selected from the null space. In this case we can not employ the solution suggested by null space selection and Problem (\ref{eq:w_update_reg1}) must be solved for each time. Alg. \ref{alg:centralized} summarizes the steps of the proposed algorithm. First beamforming vectors are initialized randomly and the corresponding propagation matrix $\boldsymbol{\Gamma}$ is constructed. Then, for each time slot measurements are sensed and  the joint $\boldsymbol{x}$ estimation and beamforming problem is solved. 
%
% ---------------------------------------------------------------
%
% ---------------------------------------------------------------
%********************************************************************
\section{Distributed Implementation} \label{sec:implementation}
\begin{figure}[b]
\centering
\includegraphics[width=1.7in,height=1.3in]{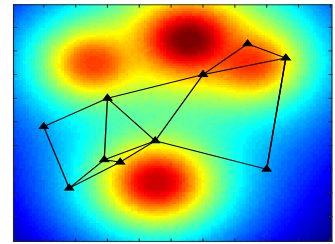}
\caption{\small{A network of $10$ SUs. They collaboratively estimate the location and power of PUs. Each SU is linked to only a subset of SUs which is determined by the connectivity graph.}}
\label{fig:distributed}
\end{figure}

In some practical scenarios there is no FC to aggregate measurements and to perform the centralized algorithm. Still, we can leverage on limited communication between neighboring SUs to enable collaborative spectrum sensing in a distributed fashion, using the beam steering capabilities of the antennas. 
%However, there is a link between neighbor SUs and they can communicate. 
Each sensor processes its received signal separately. As neighboring SUs communicate over inter-sensor links (governed by the connectivity graphs), each sensor learns and collects signals of its neighbors.  Consequently SUs can reach a consensus over the result of spectrum sensing and their estimates of propagation power vector $\boldsymbol{x}$. Fig. \ref{fig:distributed} shows a simple distributed network in which $10$ SUs estimate the locations and the signal powers of $4$ active PUs. 
%
% ---------------------------------------------------------------
% ---------------------------------------------------------------
%
%
Sensor $n$ is connected to a subset of SUs which is determined by set $\mathcal{N}_n$. Distributed version of Problem \eqref{eq:break1} is proposed as the following:
% ---------------------------------------------------------------
\begin{align}
    \label{eq:distributed1}
 \hat{\boldsymbol{x}}^{(n)} (t)   =   \underset{\boldsymbol{x}}{\text{argmin}}& \|\boldsymbol{y}^{(n)} (t)-\boldsymbol{\Gamma}^{(n)}\boldsymbol{x}-\boldsymbol{m}_v\|_2^2 \nonumber \\&+\lambda \|\boldsymbol{x}\|_1 + {\alpha} \!\! \sum_{i\in \mathcal{N}_n} \!\! \| \hat{\boldsymbol{x}}^{(i)}(t\!-\!1)-\boldsymbol{x}\|_2^2.
\end{align}
% ---------------------------------------------------------------
In this problem $\hat{\boldsymbol{x}}^{(n)}$ is the obtained estimation of  propagation power vector at the $n^{\text{th}}$ SU. Vector $\boldsymbol{y}^{(n)}$ in the collection of sensing measurements from $\mathcal{N}_n$ sensors for recent $B$ time slots and $\boldsymbol{\Gamma}^{(n)}$ is the corresponding propagation matrix. The last term of optimization applies consensus among connected SUs for each estimation. Parameter $\alpha$ regularizes the impact of the consensus term in the main cost function. Alg. \ref{alg:dist} shows the steps of the distributed implementation of our proposed algorithm.
%
% ---------------------------------------------------------------%
\begin{algorithm}[t]
\caption{Distributed spectrum sensing and beamforming.}
\textbf{Input}: Location of SUs, Neighborhood of SUs and RSS.\vspace{1mm}\\
\textbf{Output}: $\boldsymbol{x}^{(n)}(t)$ (estimated $\boldsymbol{x}$ at sensor $n$ and time $t$) \vspace{1mm}\\
\vspace{1.6mm}
\hspace{-1mm} \small{1:}  Randomly initiate of $\boldsymbol{w}_n(t)\;\forall n$.\\
\vspace{1.6mm}
\hspace{-1mm} \small{2:}  Construct $\boldsymbol{\Gamma}^{(n)}$ using Eq. (\ref{eq:RSS}).\\
\vspace{1.2mm}
\hspace{-1mm}\textbf{for} a new time slot $t$\\
\vspace{1.2mm}
$\qquad$ \textbf{for} sensor  $n$\\
 \vspace{1mm}
  \small{3:}  $\quad\qquad$ Collect vector $\boldsymbol{y}^{(n)}$ from $\mathcal{N}_n$ in recent $B$ time slots\\ \vspace{2mm}
    \small{4:} $\quad\qquad$ Construct matrix $\boldsymbol{\Gamma}^{(n)}$ for sensors indicated by $\mathcal{N}_n$ \\ \vspace{2mm}
    \small{5:}  $\quad\qquad$ $\hat{\boldsymbol{x}}^{(n)} \leftarrow$ Solve Problem (\ref{eq:distributed1})\\\vspace{2mm}
  $\qquad\qquad$ \textbf{if} $t$ {\it{mod}} $B=0$\\ \vspace{2mm}
   \small{6:}   $~\qquad\qquad$ $\hat{\boldsymbol{w}}_n (t\!+\!1)\leftarrow$ Eq. (\ref{eq:w_track}).\\\vspace{2mm}
  $~~~\quad\qquad$ \textbf{else} \\\vspace{2mm}
  \small{7:}   $~\qquad\qquad$ Construct matrix $\boldsymbol{U}^{(n)}$ for $\mathcal{N}_n$ in recent $B$ time slots.\\\vspace{2mm}
   \small{8:}   $~\qquad\qquad$ $\hat{\boldsymbol{w}}_n (t\!+\!1)\leftarrow$ Problem (\ref{eq:w_update_reg1}) using $\boldsymbol{U}^{(n)}$ and Eq. (\ref{eq:w_update2}).\\
   $~~~\quad\quad\qquad$\textbf{end} \\
$~\quad\qquad$\textbf{end}\\ 
\textbf{end}
\label{alg:dist}
\end{algorithm}
% ---------------------------------------------------------------%
%
%
\par Beamforming optimization is similar to the centralized solution. However, each SU only has access to the scanned beams of itself and its neighbors to construct  matrix $\boldsymbol{U}^B$ in problem (\ref{eq:w_update_reg1}). We define matrix $\boldsymbol{U}^{(n)}(t)$ for each time which corresponds to the already scanned beams in recent $B$ time slots via sensor $n$ and its neighbors. Mathematically, the set of rows of matrix $\boldsymbol{U}^{(n)}(t)$ can be expressed by,
% ---------------------------------------------------------------
\begin{align}
\{ \boldsymbol{u}_j(\tau)\}\;\; \small{\text{where,}}&\;\; \boldsymbol{u}_j(\tau)=\boldsymbol{w}_j^H(\tau) \boldsymbol{A}_j \nonumber\\ &\forall j\in \mathcal{N}_n \; \small{\text{and}} \; \tau=t-B+1,\cdots,t-1.\nonumber
\end{align}
% ---------------------------------------------------------------
The goal is to add a new row such that the criterion in Problem (\ref{eq:w_update_reg1}) is optimized. 
% ---------------------------------------------------------------
%****************************************************************
%****************************************************************
% ---------------------------------------------------------------
\section{Radio Cartography} \label{sec:cartography}
\begin{figure}[!t] 
\centering
\includegraphics[width=3.45 in,height=2.55in]{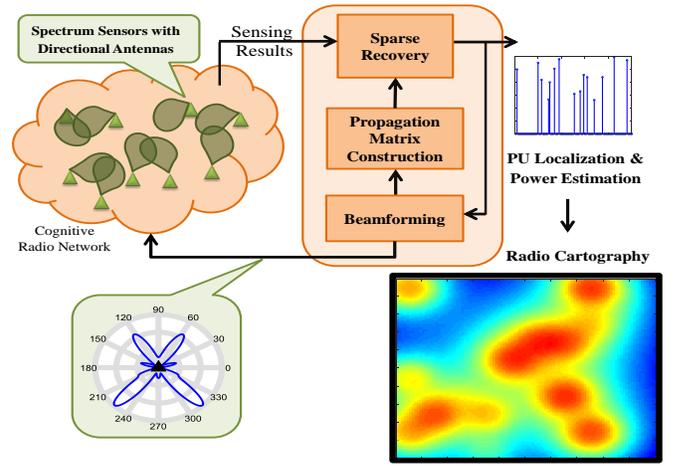}
\vspace{-4mm}
\caption{\small{Block diagram of the dynamic spectrum sensing scheme which uses our proposed localization and beamforming algorithm. }}
\label{fig:block_diag}
\vspace{0mm}
\end{figure}

An important side product of estimating $\boldsymbol{x}(t)$ in Alg. \ref{alg:dist} and Alg. \ref{alg:centralized} is generating the RF power map for any arbitrary point in the network. This is also referred to as radio cartography. Consider a point $\boldsymbol{g}$ with coordinate $(g_1, g_2)$ in the network. To estimate the RSS at point $(g_1, g_2)$, it is sufficient to consider the estimated sources at their corresponding locations and apply the underlying propagation model. 
Propagation pattern of sources is considered to be omni-directional. The estimated RSS at location $\boldsymbol{g}$ and time instant  $t$ can be interpolated as follows:
% ---------------------------------------------------------------
\begin{equation} \label{eq:rss_g}
    RSS(\boldsymbol{g})=\sum_{p=1}^P \frac{\hat{x}_p(t)}{R_{g,p}^{\eta}}.
\end{equation}
% ---------------------------------------------------------------
where $R_{g,p}$ is the distance between location $g$ and the $p$-th PU. Fig. \ref{fig:block_diag} shows the block diagram of dynamic spectrum cartography via our proposed localization and beamforming algorithm.
% ---------------------------------------------------------------
% ---------------------------------------------------------------
% ******************************************************
% ******************************************************
\vspace{+3mm}
\section{Experimental Results} 
\label{sec:experiments}

\begin{figure}[!t]
\vspace{-2mm}
\centering
\begin{subfigure}{0.23\textwidth}
\centering
\includegraphics[width=1.45in]{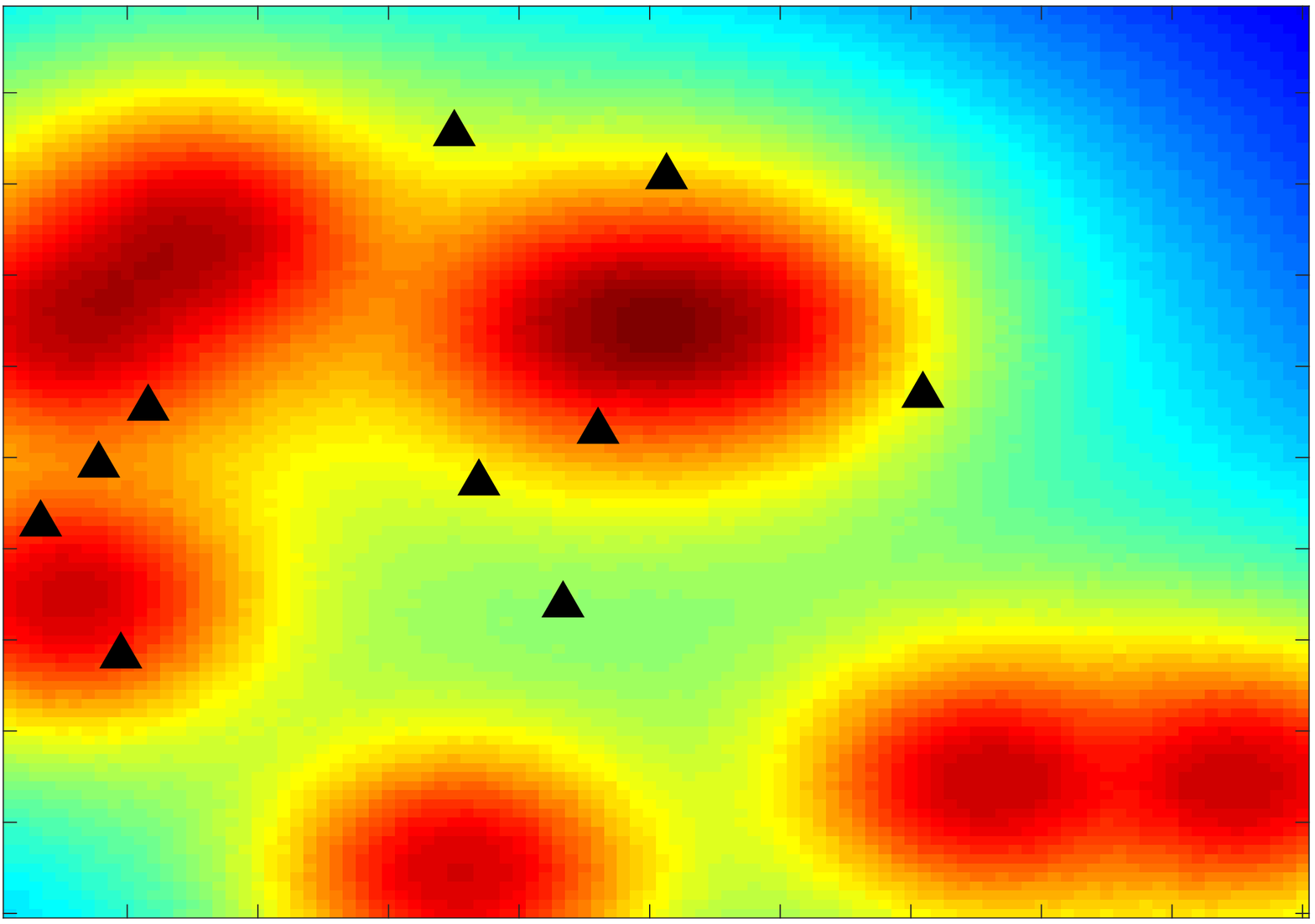}
\vspace{-1mm}
\footnotesize{\caption{}}

\end{subfigure}
\begin{subfigure}{0.23\textwidth}
\includegraphics[width=1.45 in]{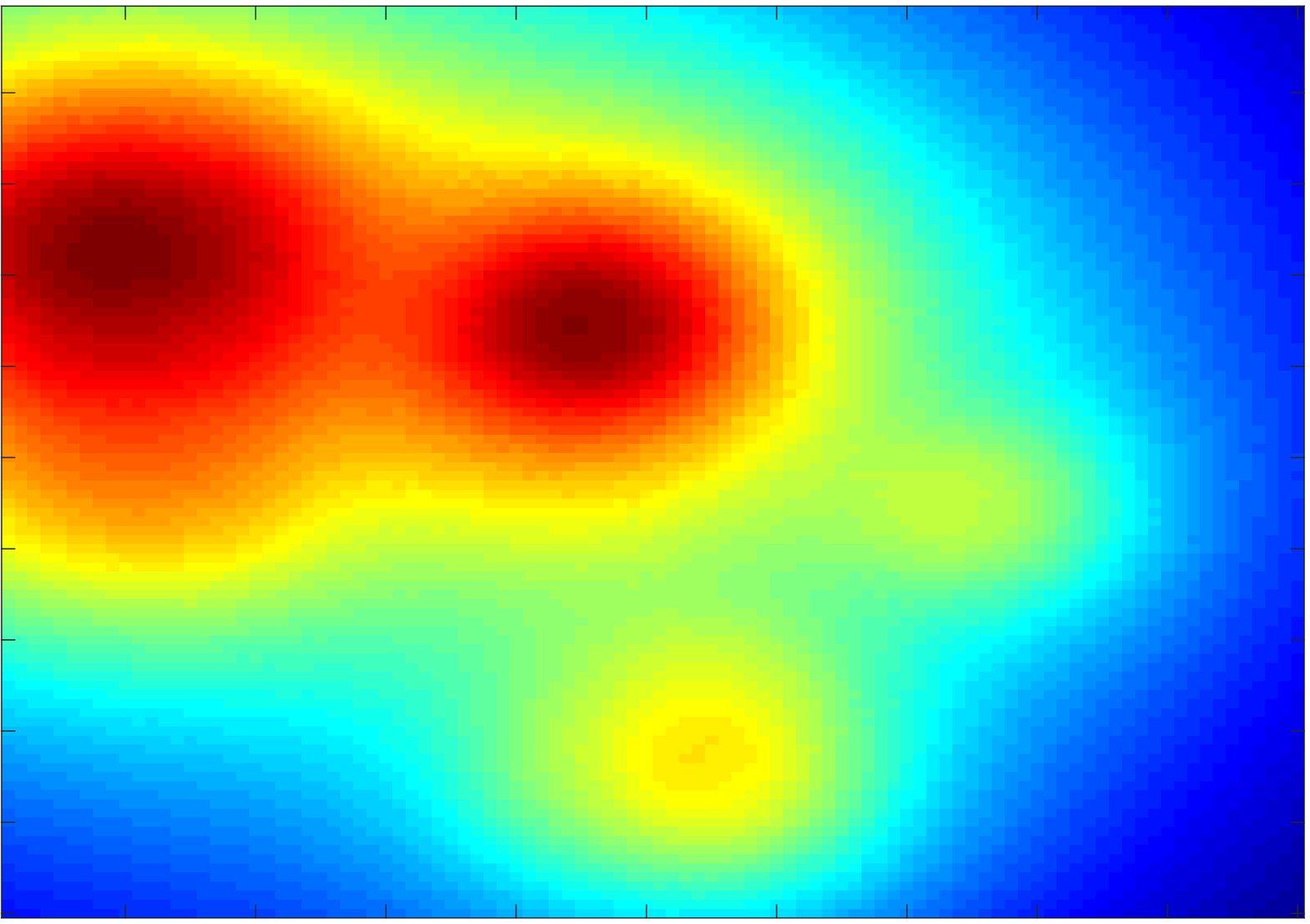}
\centering
\vspace{-1mm}
\small{\caption{}}
\end{subfigure}
\begin{subfigure}{0.23\textwidth}
\includegraphics[width=1.45 in]{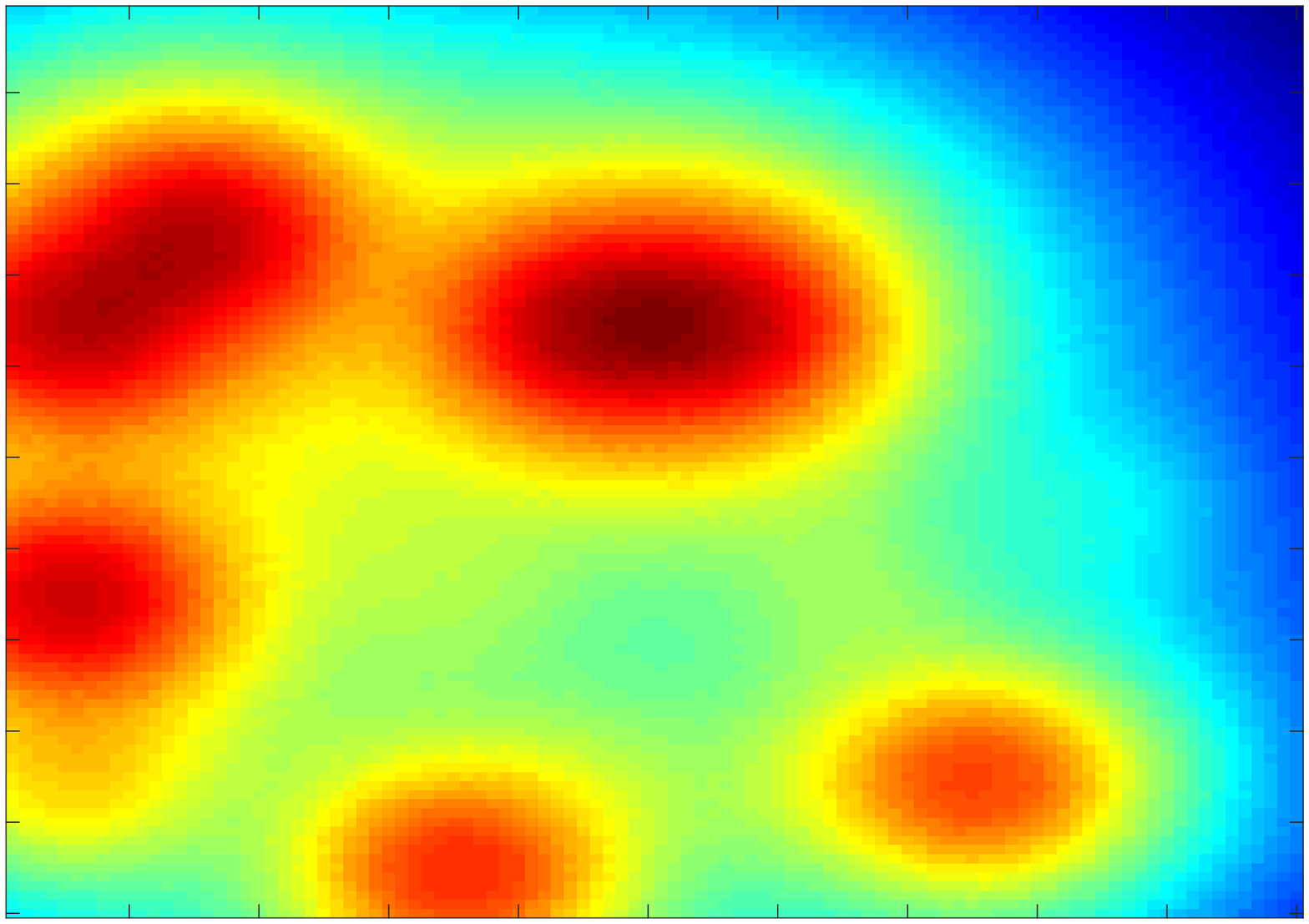}
\centering
\vspace{-1mm}
\small{\caption{}}
\end{subfigure}
\begin{subfigure}{0.23\textwidth}
\includegraphics[width=1.45 in]{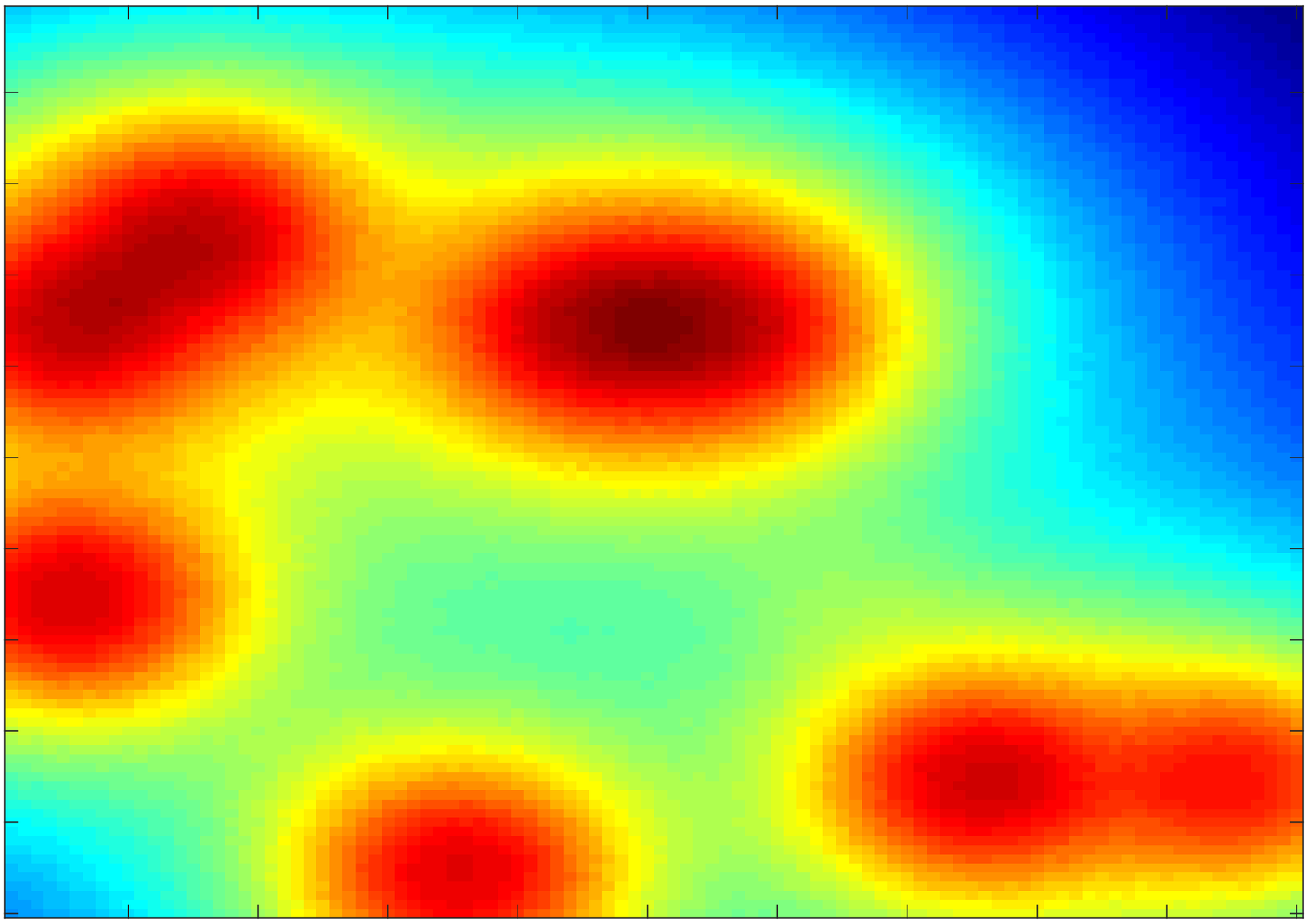}
\centering
\vspace{-1mm}
\small{\caption{}}
\end{subfigure}
\caption{\small{Comparison of estimated power map with the ground truth. (a) Ground truth that synthesized data are generated accordingly. (b) The estimated map using omni-directional SUs \cite{Bazerque10Distributed}. (c) The estimated map utilizing directional antennas with randomly chosen initial patterns. (d)  Retrieved spectrum map using the proposed algorithm after $20$ time slots.}}
\label{fig:cpmpare}
\vspace{-3mm}
\end{figure}

In this section, experiment results for synthetic data in a cognitive radio network are presented. Simulations are performed using CVX toolbox in Matlab.
The area of interest is $100$m $\times 100$m. The locations of PUs on this grid are unknown and SUs are placed randomly on this grid. 
Fig. \ref{fig:cpmpare} shows the impact of directional antennas on the localization accuracy. There are $8$ active PUs in the network and 10 SUs are sensing the environment. In this figure there is no noise on the RSS and our proposed method reaches almost exact reconstruction after few time slots. 
Fig. \ref{fig:cpmpare}a shows the ground truth power map and location of $10$ SUs in the network. The estimated power map using omni-directional antennas is depicted in Fig. \ref{fig:cpmpare}b.

Employing Directional antennas by randomly chosen beamforming weights provides a significant improvement compared to omni-directional antennas which is shown in Fig. \ref{fig:cpmpare}c. Our proposed method reaches exact reconstruction after $50$ time slots. Fig. \ref{fig:cpmpare}d shows the reconstructed power map using our algorithm that is identical to the ground truth. 
%
%
% ---------------------------------------------------------------
% ---------------------------------------------------------------
%
%
%
\par
\begin{figure}[!t]
\vspace{0mm}
\centering
\includegraphics[width=2.55in,height=1.65in]{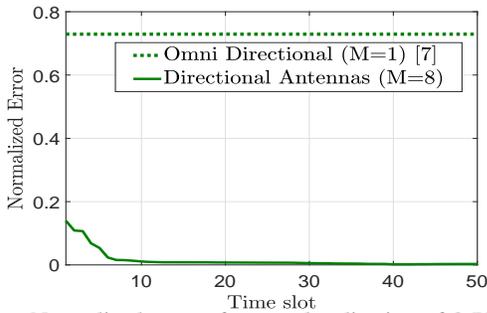}\vspace{-2mm}
\caption{\small{Normalized error of source localization of $8$ PUs. }}
\label{fig:error1}
\vspace{-4mm}
\end{figure}

\begin{figure}[!t]
\centering
\includegraphics[width=2.55in,height=1.65in]{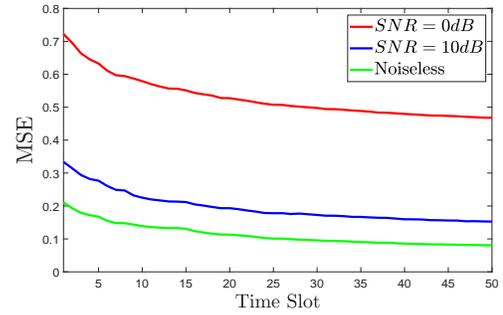}
\caption{\small{Normalized error of source localization for two SNRs and the noiseless regime over time.}}
\label{fig:error_noise}
\vspace{-1mm}
\end{figure}

\begin{figure}[!t]
\centering
\includegraphics[width=2in,height=1.5 in]{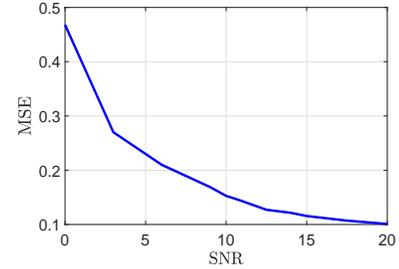}
\caption{\small{Normalized error of our centralized algorithm versus SNR.  }}
\vspace{0mm}
\label{fig:SNRs}
\vspace{-3mm}
\end{figure}

\begin{figure}[!t]
\vspace{0mm}
\begin{subfigure}{0.5\textwidth}
\centering
\includegraphics[width=2in,height=1.5 in]{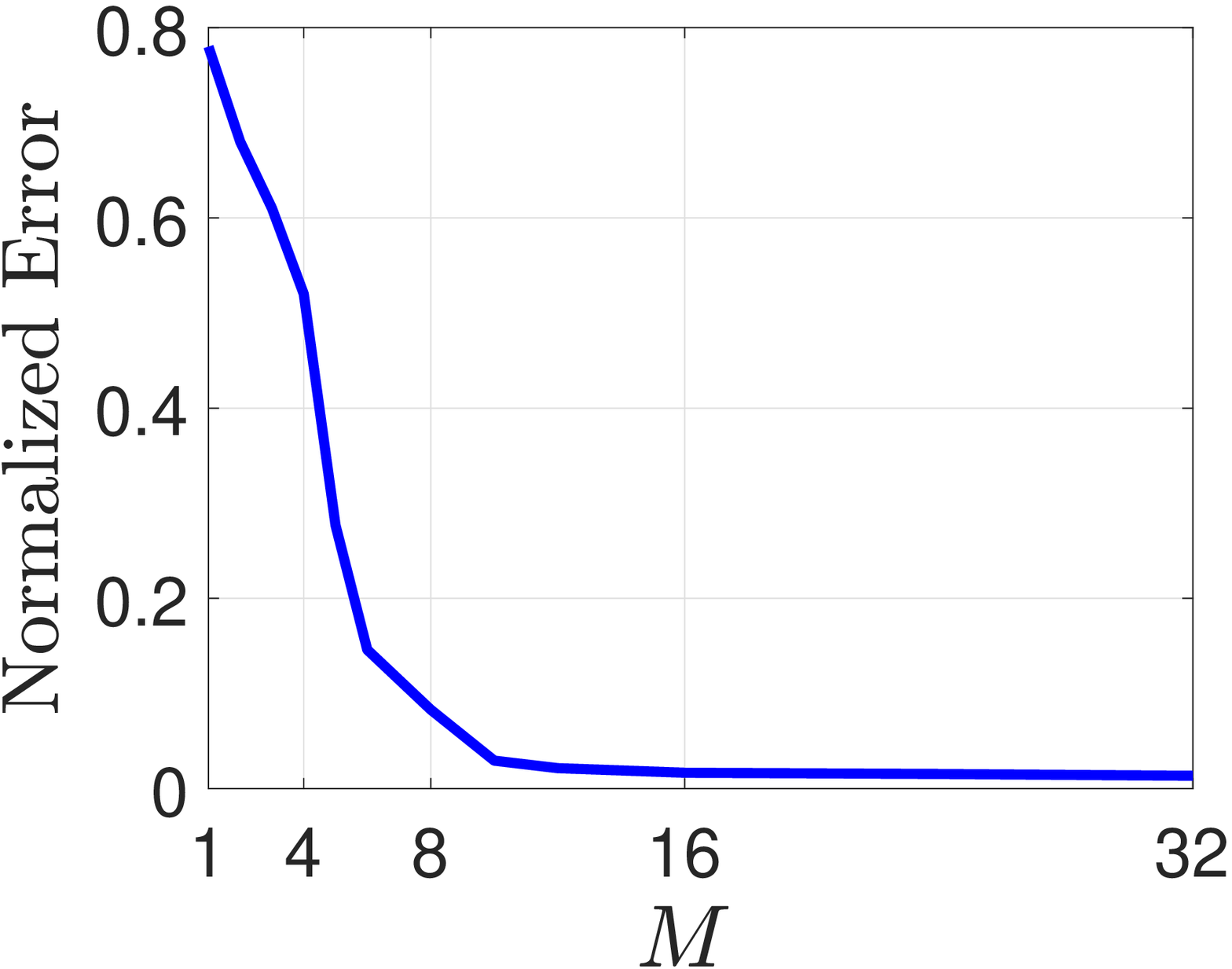}
\caption{}
\label{fig:M}
\end{subfigure}
\begin{subfigure}{0.5\textwidth}
\centering
\includegraphics[width=2in,height=1.5 in]{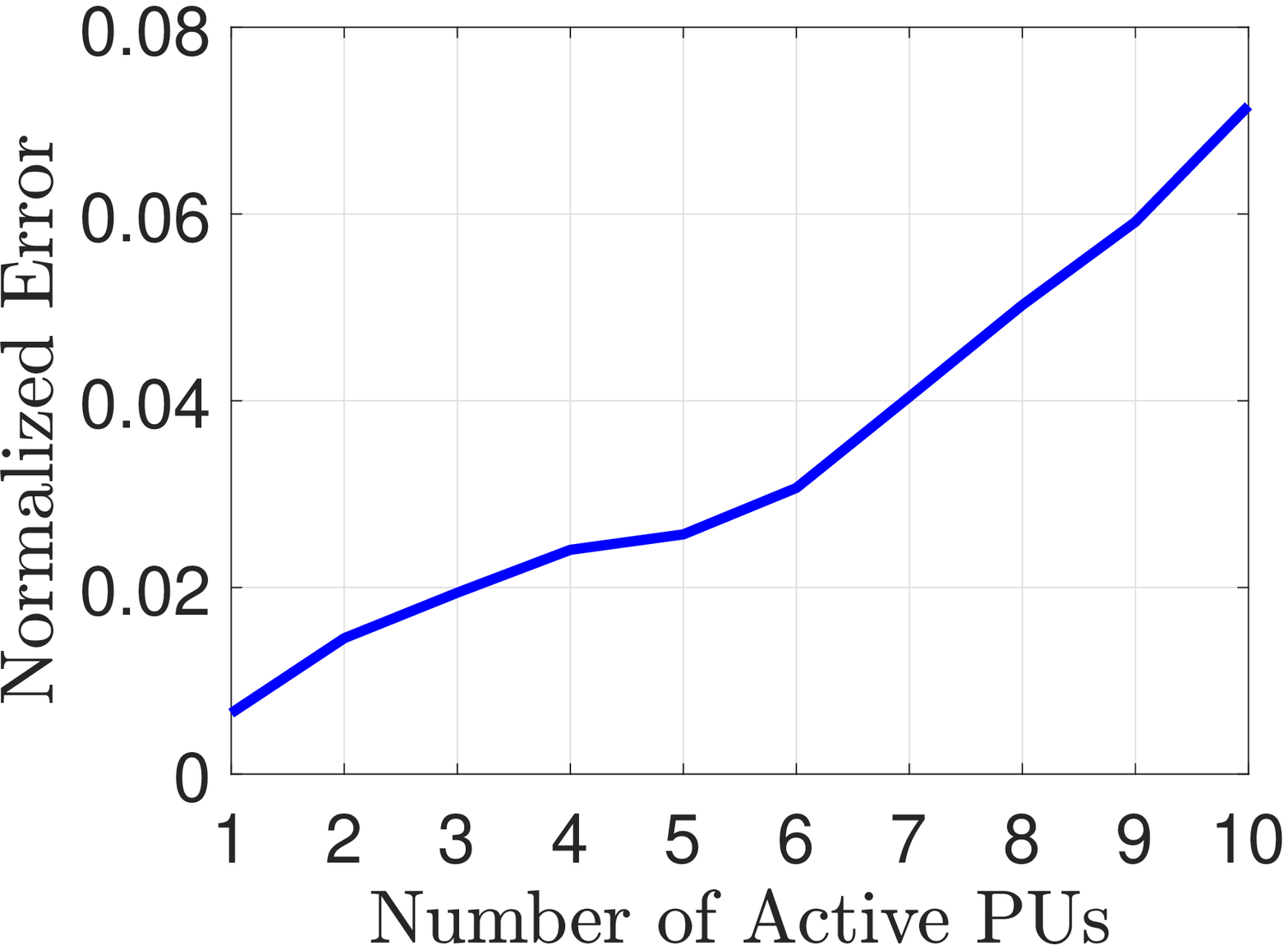}
\caption{}
\vspace{-1mm}
\label{fig:activepu}
\end{subfigure}
\caption{\small{(a) Normalized error of source localization in terms of $M$, (b) Normalized error versus number of active PUs.}}
\vspace{-4mm}
\end{figure}

 Fig. \ref{fig:error1} exhibits the numerical comparison corresponding to the simulation of Fig. \ref{fig:cpmpare}. This graph shows the performance of localization in terms of normalized error of estimation as follows:
% ---------------------------------------------------------------
\begin{equation}
    \small{\text{normalized error}}(t)=\frac{\|\boldsymbol{x}(t)-\hat{\boldsymbol{x}}(t)\|_2}{\|\boldsymbol{x}(t)\|_2},
\end{equation}
% ---------------------------------------------------------------
in which $\hat{\boldsymbol{x}}(t)$ is the estimated vector that contains location and power of each PU at time $t$. As it can be seen,  after $20$ time slots our algorithm retrieves the true locations for PUs. 
%
% ---------------------------------------------------------------
% ---------------------------------------------------------------
%
%
The effect of additive noise is studied in Fig. \ref{fig:error_noise} over time. As it can be seen, the performance of localization when $\text{SNR}=10$ dB is close to the noiseless regime. The noise is added according to \eqref{eq:rn}. A Monte Carlo simulation with $50$ different noise realization is performed for three level of noise. In this figure mean square error (MSE) is evaluated over time which is equal to average of normalized errors.   
%
% ---------------------------------------------------------------
% ---------------------------------------------------------------
%
%
Fig. \ref{fig:SNRs} indicates the normalized error after $20$ time slots of the algorithm versus different SNRs. 
%
% ---------------------------------------------------------------
% ---------------------------------------------------------------
%
%
In Fig. \ref{fig:M} we investigate the impact of the number of elements $M$ in each antenna array on the normalized error of localization. 
As this figure suggests, $M=8$ is sufficient for accurate localization within the area. Fig. \ref{fig:activepu} shows performance of localization in presence of different number of active PUs in a network with $100$ grid points and $10$ sensors each one is equipped with $8$ elements. 
%
%
% ---------------------------------------------------------------
% ---------------------------------------------------------------
%
%
\par
\begin{figure}[!t]
\centering
\vspace{-3mm}
\includegraphics[width=2.65in,height=1.85in]{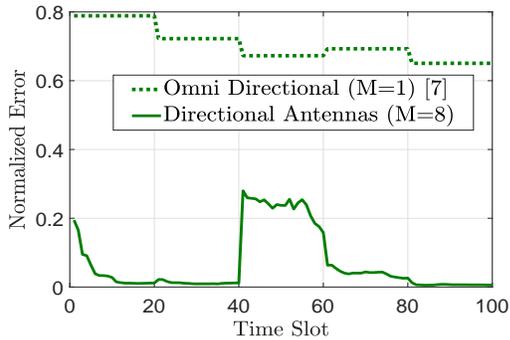}
\caption{\small{Normalized error of source localization in the presence of $8$ PUs. Each $20$ time slots the location of one PU changes. }}
\label{fig:error2}
\vspace{0mm}
\end{figure}

 Fig. \ref{fig:error2} considers the situation where there are 8 PUs in the area and the location of one PU is changing every  $20$ time slots. Each location change 
is discovered and tracked. However, the location change occurred at time slot $40$ takes a longer time to be detected and tracked. The performance of our distributed localization algorithm for detecting and tracking $8$ PUs is compared with the case where sensors have omni-directional antennas. Our algorithm tracks the location changes and refines the estimated power map over time. 
%
%
% ---------------------------------------------------------------
% ---------------------------------------------------------------
%
%
\par
\begin{figure}[!t]
\centering
\begin{subfigure}{0.23\textwidth}
\centering
\includegraphics[width=1in]{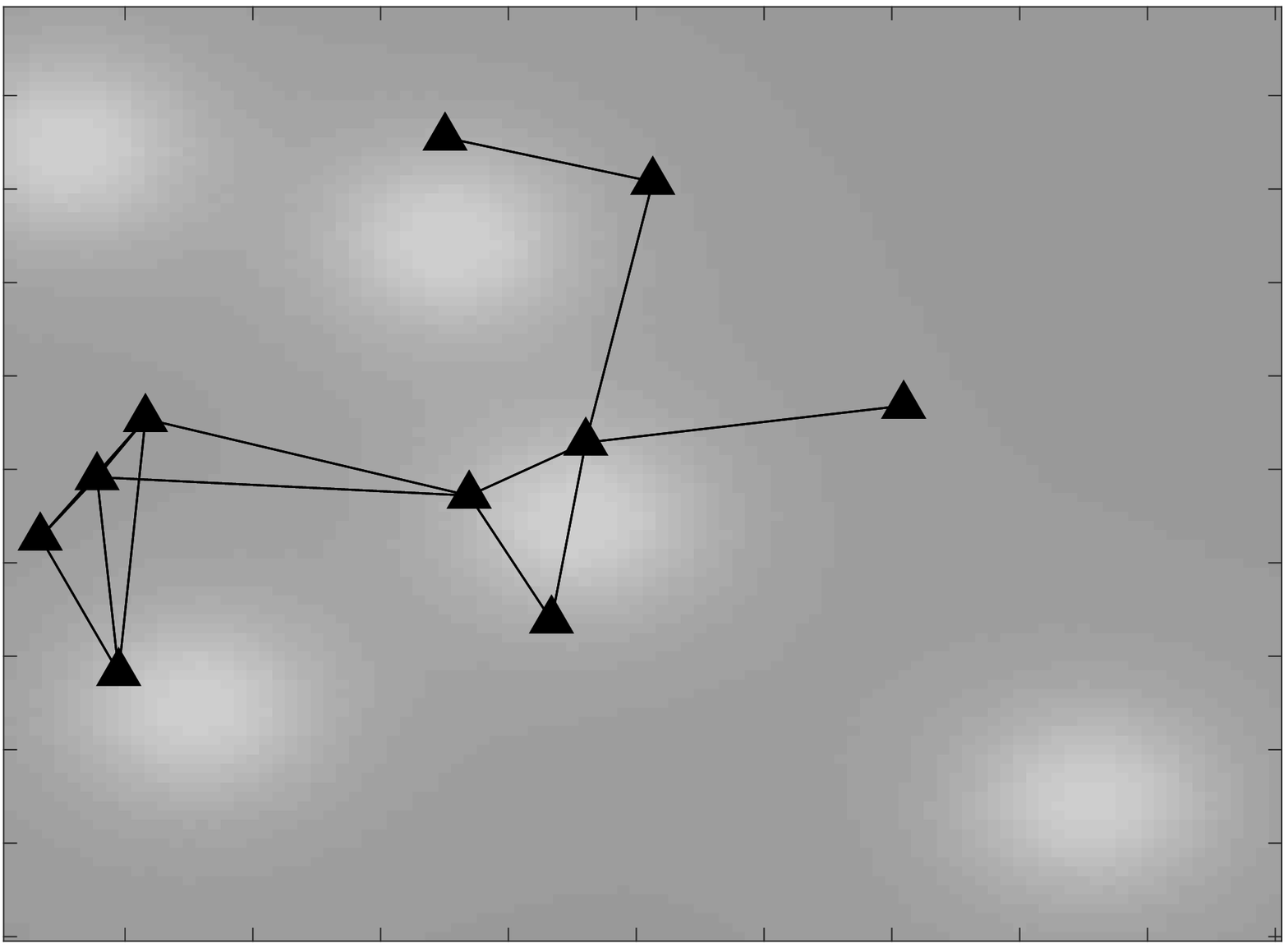}
%\vspace{-1mm}
\footnotesize{\caption{Connectivity graph 1.}}
\end{subfigure}
\begin{subfigure}{0.23\textwidth}
\includegraphics[width=1 in]{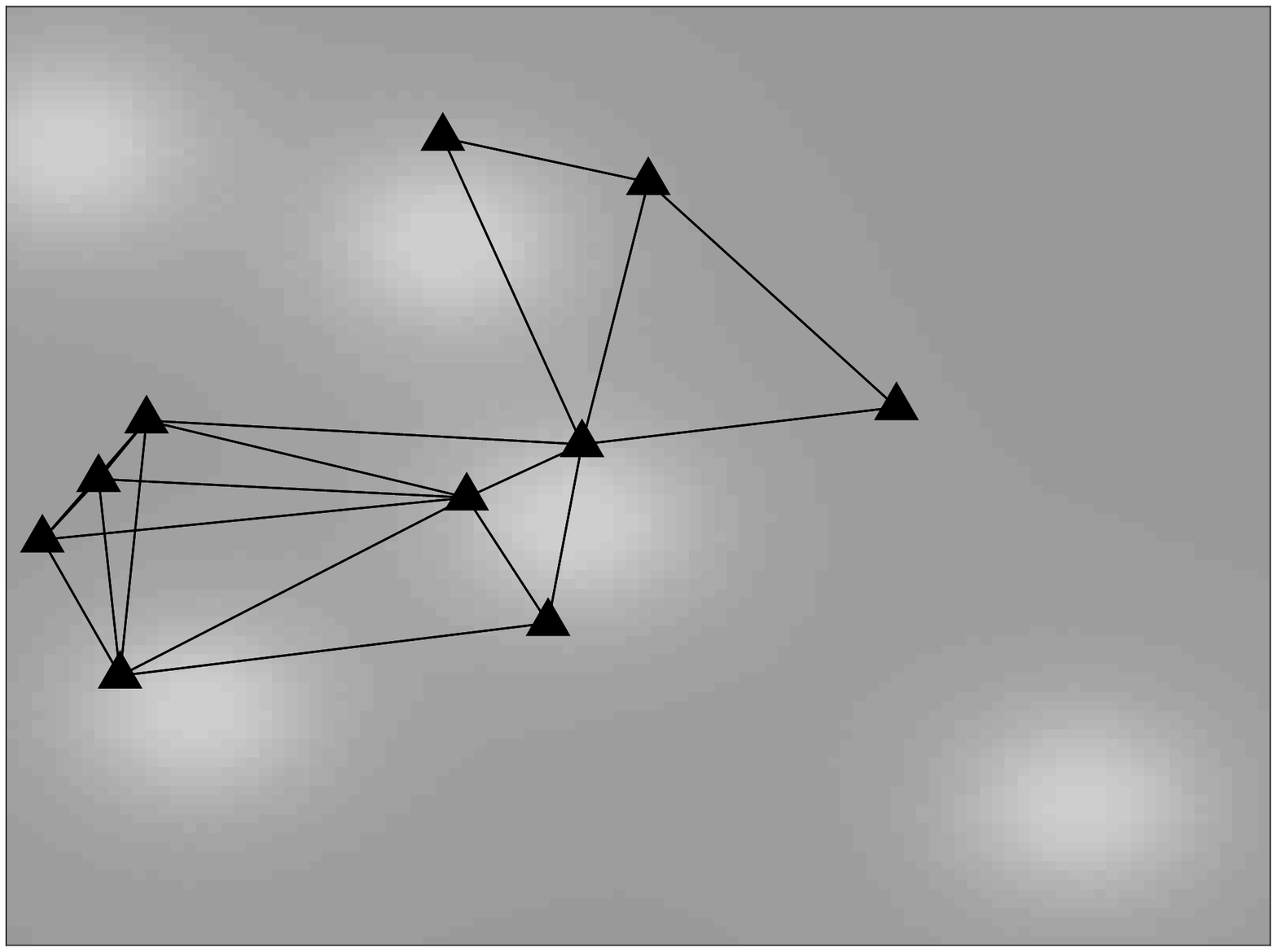}
\centering
\vspace{-1mm}
\small{\caption{Connectivity graph 2.}}
\end{subfigure}
\begin{subfigure}{0.4\textwidth}
\includegraphics[width=2.5 in]{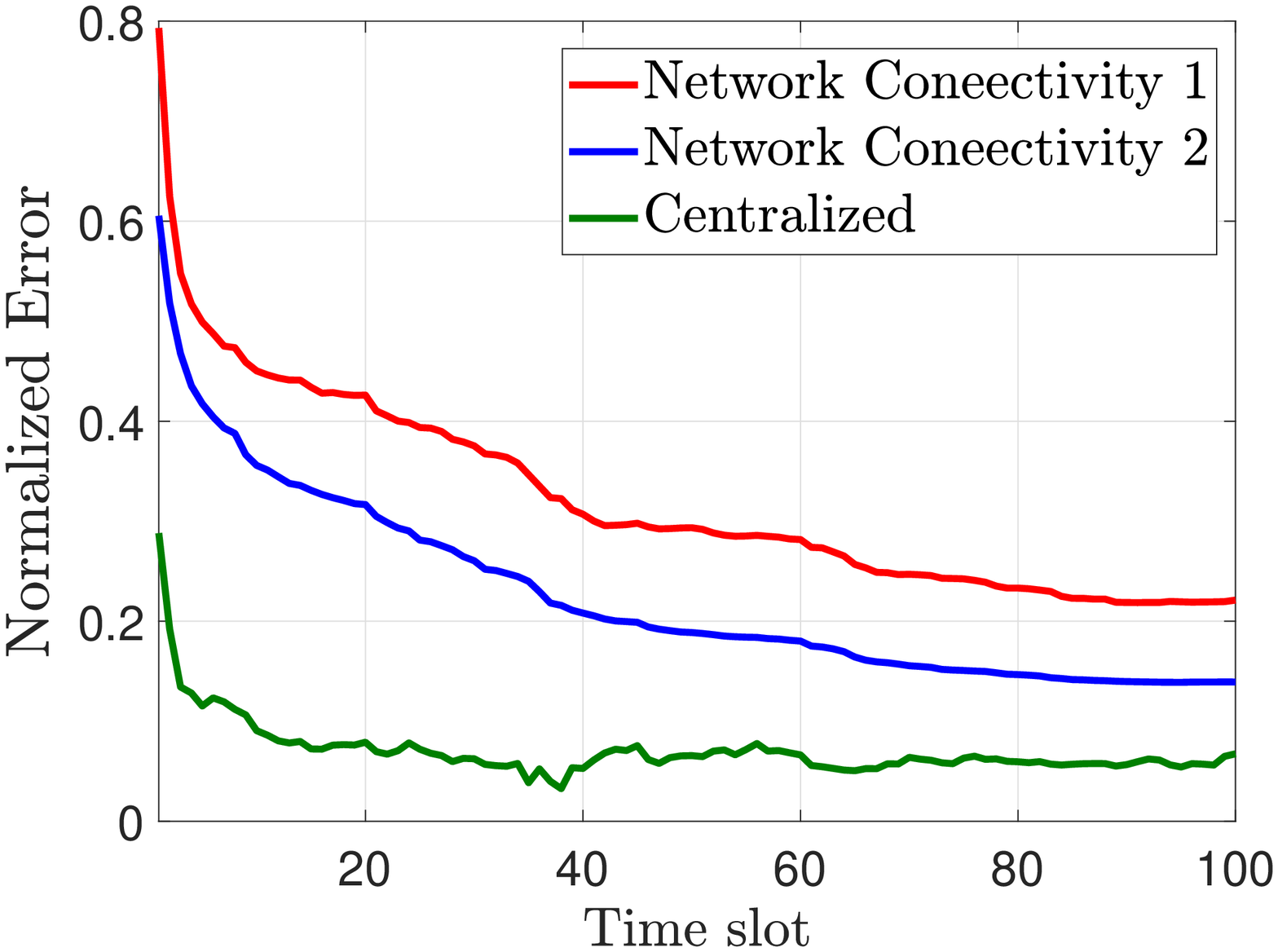}
\centering
%\vspace{-1mm}
\small{\caption{Normalized error for two connectivity graphs.}}
\end{subfigure}
\caption{\small{Effect of connectivity graphs on the performance of source localization.}}
\vspace{-4mm}
\label{fig:distributed_connectivity}
\end{figure}

 The connectivity of nodes plays a key role in the performance of our distributed algorithm. Fig. \ref{fig:distributed_connectivity}a and Fig. \ref{fig:distributed_connectivity}b compare two graphs of connectivity for $10$ spectrum sensors in which nodes are more connected to each other in Fig.  \ref{fig:distributed_connectivity}b. The normalized error of localization for these two distributed networks are compared with the centralized solution in Fig. \ref{fig:distributed_connectivity}c. As the network is more connected, sensors reach consensus faster. However, the centralized solution converges to the final solution in much less number of time slots.
%
%
% ---------------------------------------------------------------
% ---------------------------------------------------------------
%
%
\par
\begin{figure}[!t]
\centering
\includegraphics[width=2.5in,height=1.65 in]{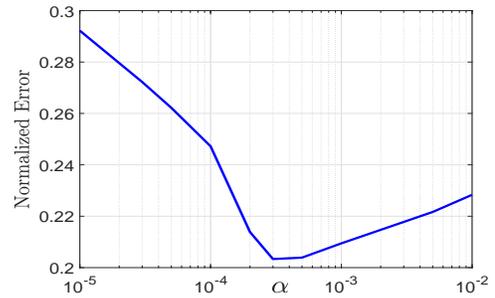}
\caption{{The impact of parameter $\alpha$ to reach consensus over time.  }}\label{fig:distributed_simul1}
\vspace{0mm}
\end{figure}

\begin{figure}[!t]
\centering
\begin{subfigure}{0.4\textwidth}
\centering
\includegraphics[width=1.5in]{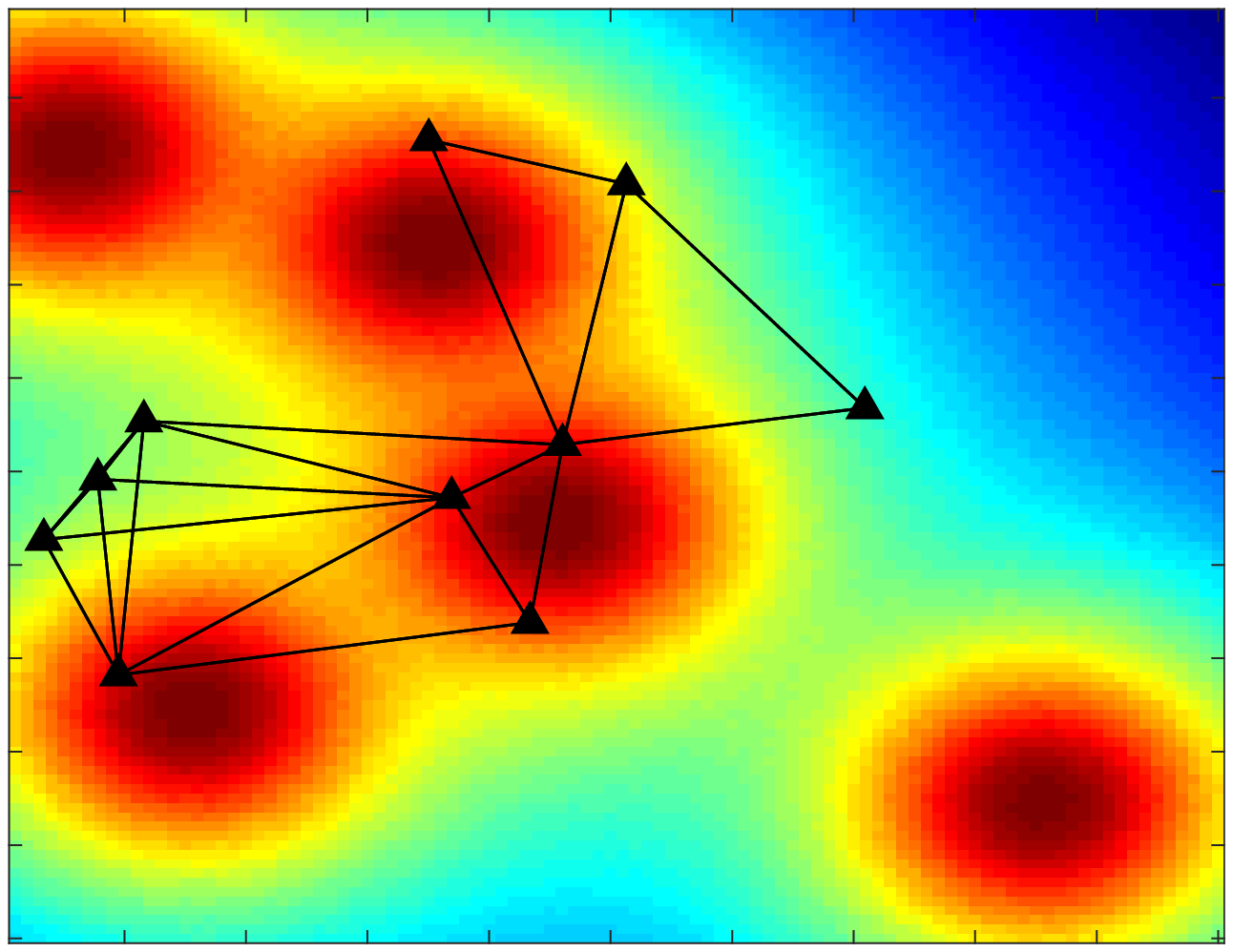}
\vspace{-1mm}
\footnotesize{\caption{The ground truth solution.}}
\end{subfigure}
\begin{subfigure}{0.23\textwidth}
\includegraphics[width=1.5 in]{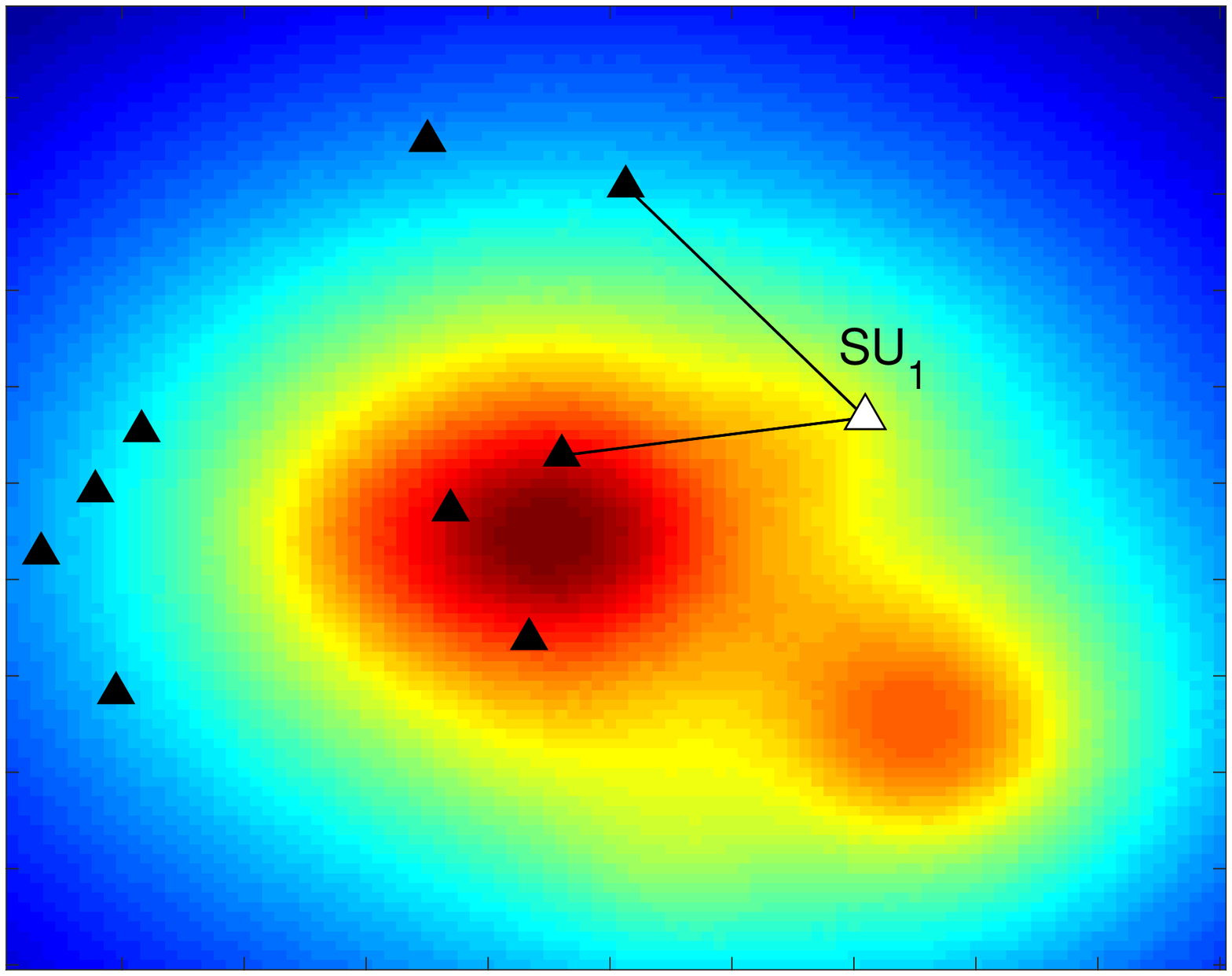}
\centering
\vspace{-1mm}
\small{\caption{Estimated power map by $\;\;\;\;\;\;\;\;\;\;\;\;$  $\text{SU}_1$ at time slot $1$.}}
\end{subfigure}
\begin{subfigure}{0.23\textwidth}
\includegraphics[width=1.5 in]{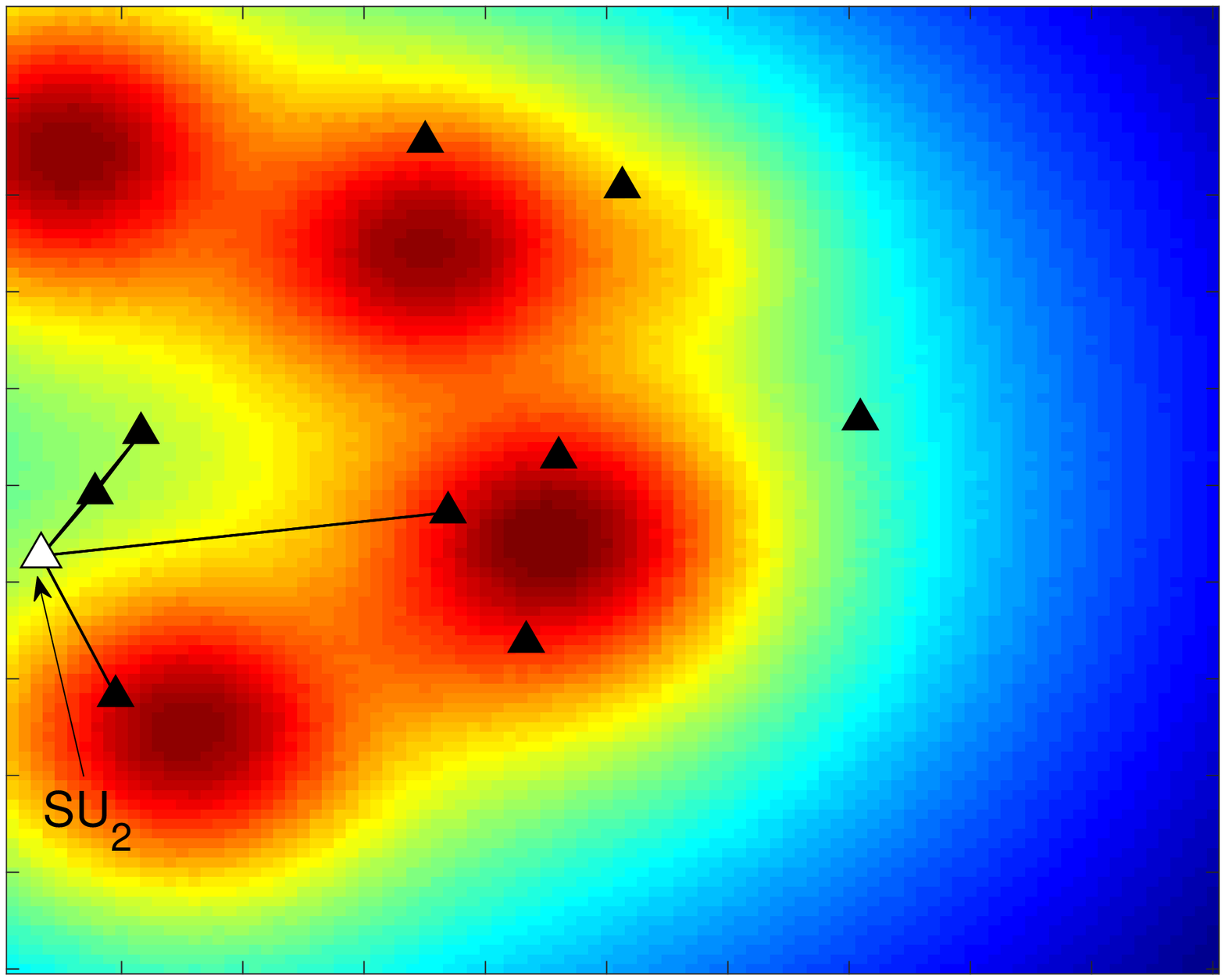}
\centering
\vspace{-1mm}
\small{\caption{Estimated power map by $\text{SU}_2$ at $\;\;\;\;\;$ time slot $1$.}}
\end{subfigure}
\begin{subfigure}{0.23\textwidth}
\includegraphics[width=1.5 in]{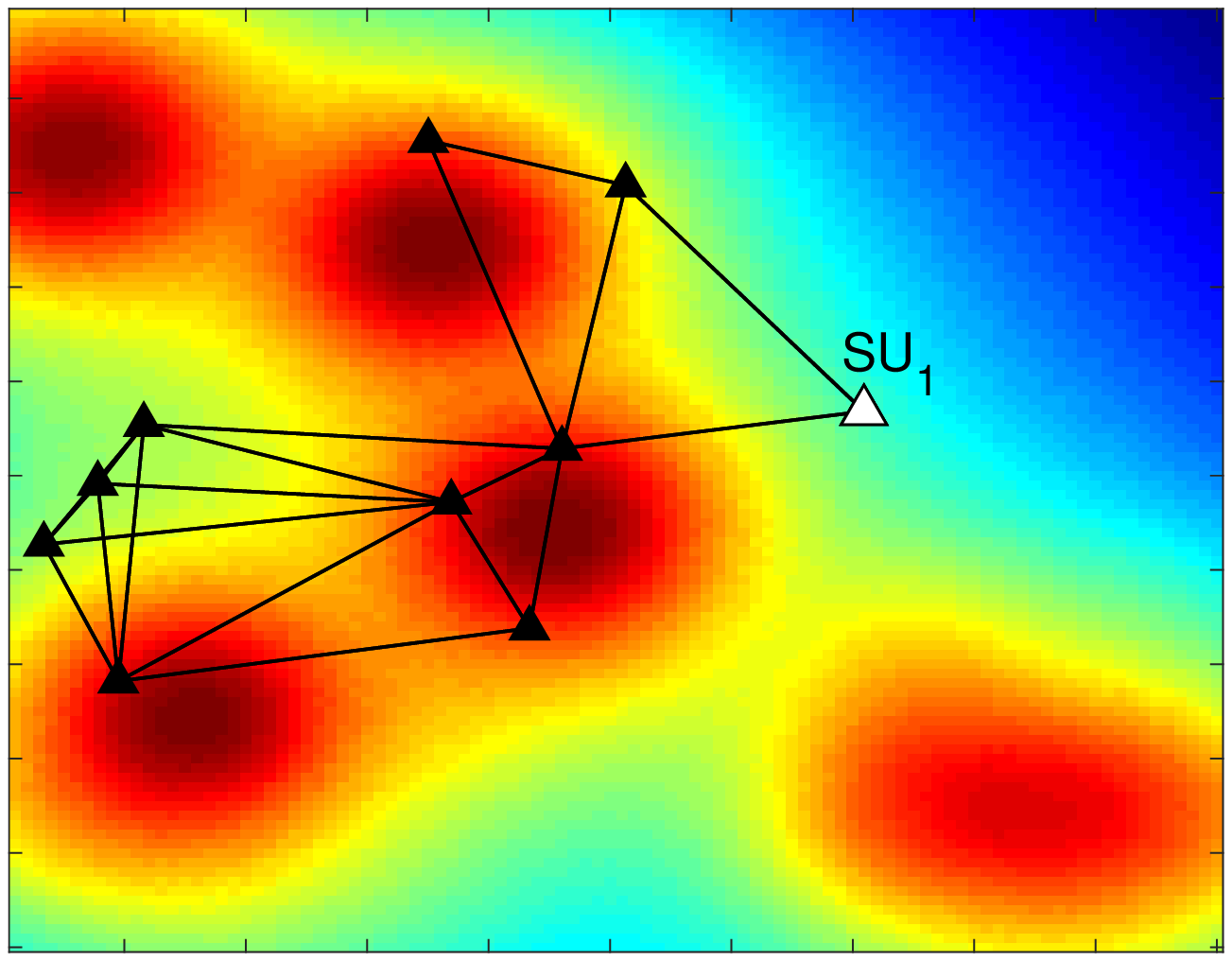}
\centering
\vspace{-1mm}
\small{\caption{Estimated power map  by $\;\;\;\;\;\;\;\;\;\;\;\;$  $\text{SU}_1$ at  time slot $100$.}}
\end{subfigure}
\begin{subfigure}{0.23\textwidth}
\includegraphics[width=1.5 in]{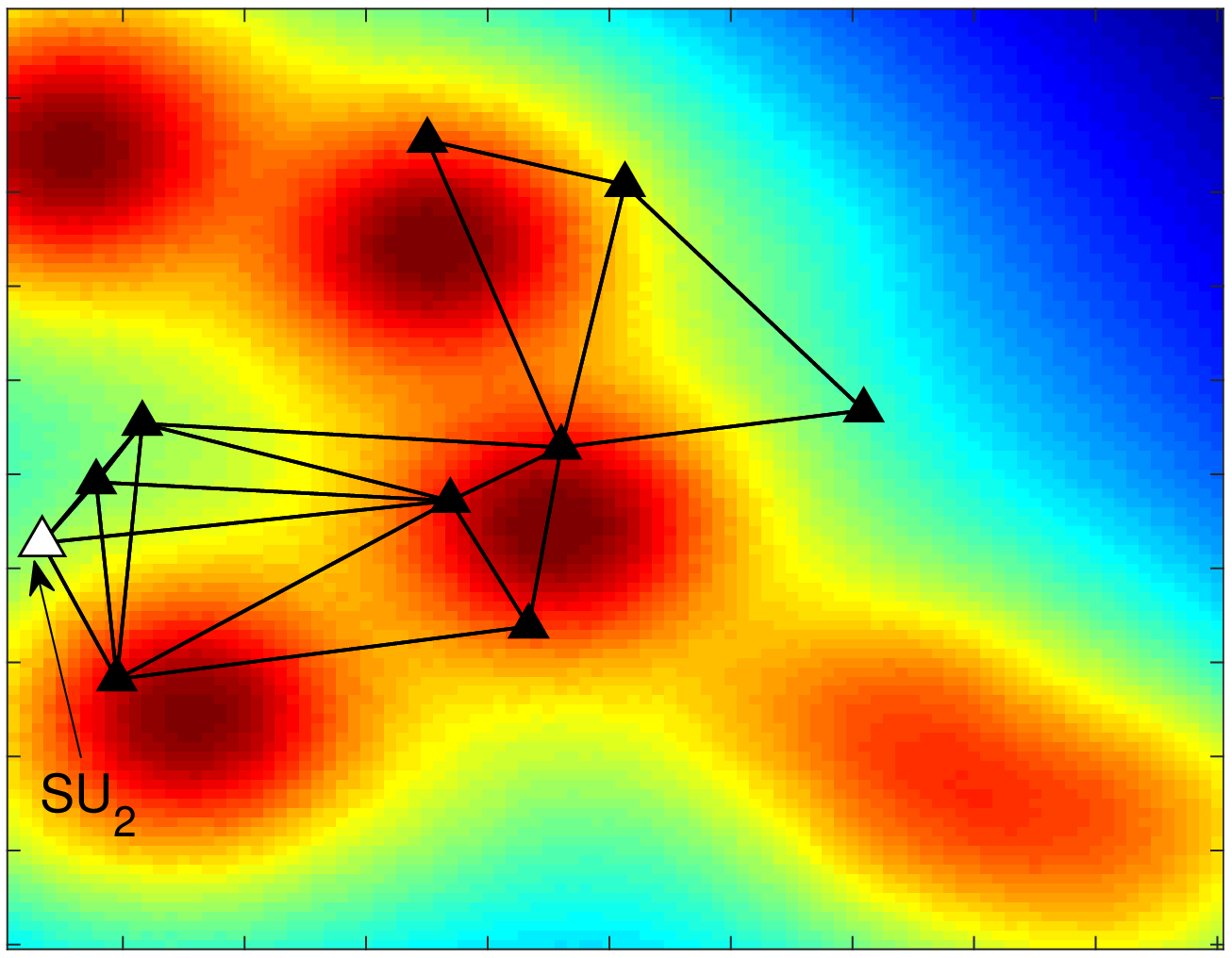}
\centering
\vspace{-1mm}
\small{\caption{Estimated power map by $\text{SU}_2$ at $\;\;\;\;\;$ time slot $100$.}}
\end{subfigure}
\caption{The accuracy of solution using the proposed distributed algorithm over time.}
\vspace{-3mm}
\label{fig:copmpare_distributed}
\end{figure}

 Fig. \ref{fig:distributed_simul1} shows performance of our proposed distributed algorithm for localization of PUs versus parameter $\alpha$ which regularizes the impact of consensus. The location of PUs are assumed to be constant over time and the distributed algorithm reaches consensus about location of PUs over time. This plot is drawn in terms of averaged normalized error of all sensors. At the first time slot different sensors may have very different estimation of PUs' locations. However, after several time slots the accuracy of their estimations increases and their solutions converge to each other. Fig. \ref{fig:copmpare_distributed} compares the estimated power map of two distant SUs. The ground truth power map is shown and the estimated power map using SU$_1$ and SU$_2$ are compared at time slots $1$ and  $100$. At time slot $100$, all sensors have used information of all other sensors and a consensus has been reached on the PUs' locations estimates.
%
%
% ---------------------------------------------------------------
% ---------------------------------------------------------------
%
%
% ---------------------------------------------------------------
% ---------------------------------------------------------------

%*****************************************************************
\vspace{0mm}
\section{Conclusion}
The impact of directional antennas on the performance of source localization and tracking is studied. Omni-directional antennas provides redundant measurements over time. While, directional antennas are able to capture a set of incoherent projections from the underlying spectrum activity pattern. It improves the accuracy of source localization significantly. As the number of elements in an ULA increase, the quality of power spectrum map increases. However, employing $8$ or $16$ elements in each ULA is sufficient for practical scenarios. The proposed localization is used in a dynamic spectrum cartography system. Moreover, the distributed implementation of our proposed method is presented and evaluated.

%***************************************************************
\balance
\small{
\bibliographystyle{IEEEtran}
\bibliography{RefFile}
}
\end{document}